\documentclass{JAIS}

\journal{JAIS-ID}
\vol{2024}

\received{xx January 2018}
\published{xx March 2018}

\def\be{\begin{equation}}
\def\ee{\end{equation}}
\def\bea{\begin{eqnarray}}
\def\eea{\end{eqnarray}}

\usepackage{url}
\usepackage{caption}
\usepackage{array}
\usepackage{tabularx}
\usepackage{ragged2e}
\usepackage{lineno}
\usepackage{amsmath}
\usepackage{natbib}
\usepackage{multirow}
\usepackage{booktabs}
\usepackage{graphicx}
\usepackage{subcaption}
\biboptions{sort&compress}
\usepackage{amsmath}
\articletype{Review Articles}

\begin{document}

\title{Axion-like Particle Detection in Alkali-Noble-Gas Haloscopes}
\author{Xiaofei Huang,\auno{1,2,3} Xiaolin Ma,\auno{4} Lei Cong,\auno{5,6} Wei Ji,\auno{5,6} Jia Liu,\auno{4,7} Wei Quan,\auno{1,2,3} and Kai Wei\auno{1,2,3}}
\address{$^1$School of Instrumentation Science and Opto-electronics Engineering, Beihang University, Beijing 100191, China}
\address{$^2$Hangzhou Innovation Institute, Beihang University, Hangzhou 310051, China}
\address{$^3$Hefei National Laboratory, Hefei 230088, China}
\address{$^4$School of Physics and State Key Laboratory of Nuclear Physics and Technology, Peking University, Beijing 100871, China}
\address{$^5$Johannes Gutenberg University, Mainz 55128, Germany}
\address{$^6$Helmholtz-Institut, GSI Helmholtzzentrum fur Schwerionenforschung, Mainz 55128, Germany}
\address{$^7$Center for High Energy Physics, Peking University, Beijing 100871, China
\\
\vspace{0.2cm}
Corresponding author: Kai Wei\\
Email address: weikai@buaa.edu.cn
}

\begin{abstract}
Revealing the essence of dark matter (DM) and dark energy is essential for understanding our universe. Ultralight (rest energy $<$10\,eV) bosonic particles, including pseudoscalar axions and axion-like particles (ALPs) have emerged among leading candidates to explain the composition of DM and searching for them has become an important part of precision-measurement science. Ultrahigh-sensitivity alkali-noble-gas based comagnetometers and magnetometers are being used as powerful haloscopes, i.e., devices designed to search for DM present in the galactic halo. A broad variety of such devices include clock-comparison comagnetometers, self-compensating comagnetometers, hybrid-spin-resonance magnetometer, spin-exchange-relaxation-free magnetometers, nuclear magnetic-resonance magnetometers, Floquet magnetometers, masers, as well as devices like the cosmic axion spin-precession experiment (CASPEr) using liquid $^{129}$Xe, prepolarized via spin-exchange optical pumping with rubidium atoms. The combination of alkali metal and noble gas allows one to take the best advantage of the complementary properties of the two spin systems.
This review summarizes the operational principles, experimental setups and the successful explorations of new physics using these haloscopes. Additionally, some limiting factors are pointed out for further improvement.

\end{abstract}

\maketitle


\section{Introduction}
Although the standard model of particle physics has presented its power in innumerable phenomena, there is strong motivation to search for beyond-standard-model (BSM) physics. Various astrophysical phenomena indicate the existence of DM via its gravitational interactions. The accelerated expansion of the universe is also explained well by dark energy. However, neither DM nor dark energy has been directly observed in the laboratory \cite{adams2023axion,Aybas2021,Irastorza2021}. 
Although the large-scale collider experiments are carried out to explore dark-matter particles that could be produced in TeV-energy-scale collisions, precision measurements in atomic, molecular, and optical (AMO) physic provide access to a complementary dark-matter parameter space \cite{BAUER201516} with ultrahigh energy resolutions that are inaccessible by the traditional particle detectors. 

There are a number of major developments in recent years in the field of atomic physics, including  
laser cooling and trapping (Nobel Prize, 1997), Bose-Einstein condensation (Nobel Prize, 2001), laser-based precision spectroscopy (Nobel Prize, 2005), and measuring and manipulation of individual quantum systems (Nobel Prize, 2012). 
Advances in the fields of photonics, atomic manipulation techniques, and multifrontier scientific explorations  have opened the door to precision measurements. With a deepening understanding of the mechanisms of light-atom interaction in quantum precision measurement, spin-based quantum sensors have offered ultrahigh sensitivity, enabling sensitive searches for new physics \cite{JacksonK2023}, for instance, dark matter, Fifth forces, and the discrete-symmetry-violating permanent electric dipole moments (EDMs).

Due to their chemical inertness and long-lived coherence, noble gases (NG) are suitable for magnetic resonance imaging, neutron spin filters, inertial rotation sensing and new physics searches. In 1960, Hughes \cite{Hughes1960} and Drever \cite{Drever1961} first used nuclear spins for fundamental physics, spearheading the development of comagnetometers. Nowadays, combined noble gas and alkali magnetometers are widely used to explore spin interactions beyond the standard model \cite{Terrano2021}. In this review, we briefly introduce the principles of axion-like particles (ALP) detection via the interaction with noble gas nuclear spins. Promising haloscopes are discussed, covering dynamic models of these devices to typical experimental setups. The recent developments in ALP searches are explored, and the potential limiting factors are analyzed. While the main focus of this brief review is haloscopes, we mention some exotic-force searches as well, especially when they use instrumentation similar to that in haloscopes. We emphasize that such setups are not actually haloscopes (devices aimed at direct detection of galactic DM). On the contrary, they search for new forces that may arise due to the exchange of virtual beyond-standard-model particles. These searches are indirect rather than direct DM searches. An important feature is that such experiments are, in a sense, more general than haloscopes because they are sensitive to exotic particles whether or not these are part of galactic DM.

\section{Axion-like Particle Models}
The Hamiltonian of new-physics interactions is given by the equation ~\cite{Terrano2021,Safronova2018}:
\begin{equation}\label{eqn-21}
   \hat{H}_{\rm{BSM}} = {\vec{\boldsymbol{\sigma}}}_{\rm{n}} \cdot {\boldsymbol{\beta}}\,,
\end{equation}
where $\hat{\boldsymbol\sigma}_{\rm{n}}=\hat{\bf{F}} /F$ is the spin moment of the nucleus, ${\boldsymbol{\beta}}$ is the effective field related to the anomalous interactions with ALPs, EDM or Fifth force and so on.
\subsection{Fifth force: ALPs as virtual particle mediating forces}
New kinds of long-range forces between both macroscopic objects could be mediated by light particles that interact with electrons or nucleons. Depending on the explicit form of interaction, the effective potential of this force could be parameterized by its fermion spin, the mediator boson spin and the relative velocity between fermions, \cite{Moody1984,Dobrescu_2006,Fadeev2019}.

ALPs, if they exist, are predicted to mediate a weak, spin-dependent macroscopic force. If ALPs couples to the spins of two particle X and Y, this dipole-dipole (spin-spin) interaction can be expressed as
\begin{equation}\label{eqn-22}{}
    {\hat V_{{\rm{dd}}}} = -\frac{{g_{\rm{p}}^{\rm{X}} g_{\rm{p}}^{\rm{Y}} {\hbar ^3}}}{{16\pi {m_{\rm{X}}}{m_{\rm{Y}}}{c}{r^3}}}\left[ {\left( {{{{\boldsymbol{ \vec{ \sigma} }}}_{\rm{X}}} \cdot {{{\boldsymbol{ \vec{ \sigma} }}}_{\rm{Y}}}} \right)\left( {1 + \frac{r}{\lambda } {+\frac{4}{3}\delta(r)^3}} \right) - 3\left( {{{{\boldsymbol{\vec{ \sigma} }}}_{\rm{X}}} \cdot {\bf{\hat r}}} \right)\left( {{{{\boldsymbol{ \vec{ \sigma} }}}_{\rm{Y}}} \cdot {\bf{\hat r}}} \right)\left( {1 + \frac{r}{\lambda } + \frac{{{r^2}}}{{3{\lambda ^2}}}} \right)} \right]{{\rm{e}}^{ - r/\lambda }}\,,
\end{equation}
where $ g_{\rm{p}}^{{\rm{X}},{\rm{Y}}}$ is the pseudoscalar coupling to standard-model fermions X or Y, $m_{{\rm{X}},{\rm{Y}}}$ is the particle mass, $\lambda=\hbar/(m_a c)$ is the force range of the boson \cite{Terrano2021}, and $\delta(r)$ is the contact term \cite{Fadeev2019} with $r$ being the distance between the interacting fermions. 

If ALPs only couple to one fermion spin, the spin-mass interaction is typically called monopole-dipole interaction, and it can be expressed as 
\begin{equation}\label{eqn-23}
    {\hat V_{{\rm{md}}}} = -\frac{g_{\rm{p}}^{\rm{X}} g_{\rm{s}}^{\rm{Y}}\hbar^2 }{{8\pi {m_{\rm{X}}}}}\left[ {\left( {{{{\boldsymbol{\vec{ \sigma} }}}_{\rm{X}}} \cdot {\bf{\hat r}}} \right)\left( {\frac{1}{{r\lambda }} + \frac{1}{{{r^2}}}} \right)} \right]{{\rm{e}}^{ - r/\lambda }}=\beta\left(r\right)\hat{{\bf{r}}}\cdot\boldsymbol{\vec{ \sigma}}_{\rm{X}}=\beta\left(r\right)\frac{\left\langle \boldsymbol{\vec{ \sigma}}_{\rm{X}}  \right\rangle }{\left\langle \bf{K}  \right\rangle }\hat{\bf{r}}\cdot\bf{K}\,,
\end{equation}
where $ g_{\rm{s}}$ is the scalar coupling strength,  $\hat{\bf{r}}$ is the length unit vector connecting the two particles, $\hbar \boldsymbol{\vec{ \sigma}}_{\rm{X}} /2$ is the particle spin, and $\bf{K}$ is the nuclear spin. This interaction can also be viewed as the coupling between the spin of particle X and an effective magnetic field generated by particle Y which is proportional to the field amplitude $\beta\left(r\right)$ and the field direction along $\hat{\bf{r}}$. The interaction is presented by the shift in the precession frequency of noble gas. 
In this case, the Hamiltonian for a noble gas is \cite{Feng2022}
\begin{equation}\label{eqn-35}
    \hat{H} = -\hbar\gamma_{\rm{n}}{\bf{K}}\cdot {\bf{B}_0} + \beta\left(r\right)\frac{\left\langle {\boldsymbol{\vec{ \sigma}}_{\rm{n}}}  \right\rangle }{\left\langle \bf{K}  \right\rangle } \hat{\bf{r}}\cdot\hat{z}{\bf{K}}\cdot\hat{z}
    = -\hbar\gamma_{\rm{n}}{\bf{K}}\cdot{\bf{B}_0}+\hbar\gamma_{\rm{n}}{\bf{K}}\cdot{\bf{B}_1}\,,
\end{equation}
where $\gamma_{\rm{n}}$ is the gyromagnetic ratio of NG nuclear spin, ${\bf{B}_0}$ is the magnetic field along $\hat{z}$ and $\displaystyle{\bf{B}_1}=\frac{\beta\left(r\right)}{\hbar \gamma_{\rm{n}}}\frac{\left\langle \boldsymbol{\vec{ \sigma}}_{\rm{n}}\right\rangle }{\left\langle \bf{K}  \right\rangle }(\hat{\bf{r}}\cdot\hat{z})\cdot\hat{z}$ is the effective field produced by the anomalous interaction.
The summary of the experimental constraints from Fifth force measurements can be found in Ref.~\cite {OHare:2020wah, AxionLimits}.

\subsection{DM: ultralight ALPs behave as bosonic DM waves}
If the DM is comprised of ultralight ALPs with a mass of $10^{-22}\,\text{eV}<m_{\rm DM}<\text{eV}$, it can be identified as ultralight bosonic DM~\cite{Ferreira:2020fam,Hui:2016ltb,Sikivie:2006ni}. Ultralight bosonic DM in the galactic halo can be represented as a classcal field because of the enormous occupation number~\cite{jackson_kimball_search_2023} with its intrinsic stochastic nature~\cite{Lee2023,Centers_2021}. Since the DM is virialized in the Milky Way with velocity dispersion $v_{\rm vir}\sim 10^{-3}\,c$, this classical field is then a superposition of many oscillating fields of different frequencies with frequency dispersion $\Delta \omega\sim \frac{1}{2}m_a v_{\rm vir}^2/\hbar$, with $m_a$ being the mass of the ultralight bosonic DM. This effect induces the characteristic coherent time $\tau_c\sim m_a^{-1} v_{\rm vir}^{-2}\hbar$ and the coherent length $l_c\sim m_a^{-1}v_{\rm vir}^{-1} \hbar$.

Under the above assumptions and after certain mathematical transformations, the form of the gradient of the axion/ALP field within the galactic halo can be derived to be~\cite{Lee2023}:
\begin{align}
& \vec{\nabla}a(t) = \sum_{\Omega_\mathbf{p}} \frac{\sqrt{\rho_\textrm{DM} f(\mathbf{p}) (\Delta p)^3}}{\omega_\mathbf{p}}
\alpha_\mathbf{p} \cos(\frac{p^0 t}{\hbar} + \Phi_\mathbf{p}) \mathbf{p},\nonumber\\
&\approx \sum_{\Omega_\mathbf{v}} \sqrt{\rho_\textrm{DM} f(\mathbf{v}) (\Delta v)^3} \, \alpha_\mathbf{v} \times \cos\left(\frac{m_a}{\hbar} \left(c^2 + \frac{1}{2}v^2\right)t + \Phi_\mathbf{v}\right)\mathbf{v}\,,
\end{align}
where $\rho_{\rm DM}=0.4~\text{GeV}/(c^2\text{cm}^3)$~\cite{de_Salas_2021} represents the local DM density, $f(\bf{v})$ denotes the DM velocity distribution function under the assumption of the Standard Halo Model(SHM)~\cite{Lee_2013}, $\omega_\mathbf{p}$ denotes the axion energy associated with momentum $\mathbf{p}$, $\alpha_\mathbf{p}$ follows standard Rayleigh distribution as a random variable, and $\Phi_\mathbf{p}$ is a random phase uniformly distributed within $\left[0,2\pi\right]$. In the second line we expand the energy of nonrelativistic axion/ALP particle as a function of its velocity $v$. The summation runs over all the infinitesimal momentum space $\Omega_\mathbf{p}$ domain. One can easily check out that $\displaystyle \frac{1}{2}\langle\left(\frac{{\rm{d}} a(t)}{{\rm{d}}(ct)}\right)^2\rangle+\frac{1}{2}\frac{m_a^2c^2}{\hbar^2}\langle a(t)^2\rangle=\rho_{\rm DM} c^2$.

The nonrelativistic Hamiltonian describing the interaction between nucleus spin and axion/ALP gradient field is~\cite{Jiang2021,Bloch2022SA,Lee2023}:
\begin{align}\label{eqn-24}
    \hat{H}_{\rm ALP} &=(\hbar c)^{3/2} g_{\rm aNN} \vec{\nabla} a \cdot {\boldsymbol{ \vec{ \sigma} }}_{\rm{n}}\, ,\nonumber\\
    &=g_{\rm aNN} \sum_{\Omega_\mathbf{v}} \sqrt{\rho_\textrm{DM} f(\mathbf{v}) (\Delta v)^3\hbar^3c^3} \, \alpha_\mathbf{v} \times \cos\left(\frac{m_a}{\hbar} \left(c^2 + \frac{1}{2}v^2\right)t + \Phi_\mathbf{v}\right)\mathbf{v}\cdot {\boldsymbol{ \vec{ \sigma} }}_n.
\end{align}
where $g_{\rm aNN}$ denoted as the dimensional coupling coefficient.  
Thus, the gradient of axion/ALP acts as a magnetic field~\cite{Jiang2021,Bloch2022SA,kimball2018overview}
\begin{align}
   {\bf{B}}_{\rm{ALP}}&\approx (\hbar c)^{3/2}g_{\rm aNN} \vec{\nabla} a/(\gamma_{\rm{n}}\hbar) \cdot \frac{\left\langle {\boldsymbol{ \vec{ \sigma} }}_{\rm{n}}  \right\rangle }{\left\langle \bf{K}  \right\rangle },\nonumber\\
   &=\frac{g_{\rm aNN}}{\hbar \gamma_{\rm{n}}}\frac{\left\langle {\boldsymbol{ \vec{ \sigma} }}_{\rm{n}}  \right\rangle }{\left\langle \bf{K}  \right\rangle } \sum_{\Omega_\mathbf{v}} \sqrt{\rho_\textrm{DM} f(\mathbf{v}) (\Delta v)^3\hbar^3c^3} \, \alpha_\mathbf{v} \times \cos\left(\frac{m_a}{\hbar} \left(c^2 + \frac{1}{2}v^2\right)t + \Phi_\mathbf{v}\right)\mathbf{v}\,,
\end{align}
oscillating around the Compton frequency $\nu_a=m_a c^2/h$, with frequency dispersion $\Delta \nu_a\sim \frac{1}{2}m_a v_{\rm vir}^2/h$. The interaction between ALPs and nuclear spins is equivalent to the interaction with an ordinary oscillating magnetic field which can be expressed as 
\begin{equation}\label{eqn-25}
    \hat{H}_{\rm{ALP}} \approx \hbar\gamma_{\rm{n}}\bf{K} \cdot {\bf{B}}_{\rm{ALP}}.
\end{equation}

However, the BSM interactions are usually much weaker compared with those with real magnetic field. Thus, haloscopes are tasked with separating tiny effects from colossal backgrounds. 

For axion/ALP topological defects like domain walls coupled with the spin of nucleons, if the domain wall passes through the Earth, the nucleon spins inside the comagnetometer will be temporarily pertubated similarly to a transient magnetic field pulse~\cite{Pospelov:2012mt,JacksonKimball:2017qgk,Afach2021}. For a wide class of domain walls, one can expect many encounters between domain wall and Earth within a year thus opening a window for terrestrial experiments such as magnetometers~\cite{Pospelov:2012mt,Derevianko:2013oaa}.  While a single magnetometer alone is sensitive enough to detect this transient effect, it is necessary to separate the signal from the false signal coming from abrupt change of comagnetometer conditions like magnetic-field spikes, laser-light-mode jumps \textit{etc.} To effectively veto the false signal-like spikes, an attractive approach is to use a globally correlated magnetometers and identify the "coincidence" events. Since the comagnetometers are separated and their backgrounds are uncorrelated~\cite{Pospelov:2012mt,JacksonKimball:2017qgk,Afach2021}, this method can enhance the signal-to-noise ratio. In general four correlated magnetometers could be employed to reconstruct the velocity normal vector of domain wall, by forming a three-dimensional coordinate system. The remained magnetometers could cross-check the signal measurement by comparing the arrival time with the predicted arrival time based on the reconstructed domain wall velocity~\cite{Pospelov:2012mt}.

For axion/ALP dark matter seaches, certain peaks may show up at spectific frequencies and can be identified as signal candidates. Since the signal persists during the measurement period, one can check this candidate frequency for different part of measurement time to distinguish persistent axion signal and transient noise peaks, as demonstrated in ~\cite{Lee2023,Wei:2023rzs}. For long-standing background peaks, one can either investigate the systematic noise of the apparatus to determine its origin, or compare the shape of the false candidate peaks with axion dark matter specific signal lineshape, as done in ~\cite{Lee2023}.
Due to the stochastic nature of axion dark matter, its signal lineshape at the frequency domain prohibits a specific pattern and could be further modulated by Earth sidereal movement~\cite{Bloch2022SA,Lee2023}, which is in stark contrast with monochromatic noise. Therefore, ~\cite{Lee2023} leverages the distinctive nature of the dark matter signal to distinguish the positive signal from false positive monochromatic noise peaks.

\section{Haloscopes}
\subsection{Self-compensating comagnetometer:}
Spin-exchange relaxation limited the performance of atomic magnetometer for a long time. In 1973, Happer and his colleagues found that this limitation could be effectively suppressed or even eliminated under the conditions of high alkali-metal (AM) density and weak magnetic field, which is named the spin-exchange-relaxation-free (SERF) regime. In 2002, Allred, Romalis and their colleagues first realized an alkali-metal magnetometer based on this regime \cite{Allred2002}. Moreover, Kornack, Romalis and their colleagues demonstrated the self-compensating (SC) regime in the same year \cite{Kornack2002}, which can cancel the slow changes in the magnetic field by the interaction between NG and AM ensembles. In 2005, an SC comagnetometer operated in the SERF regime was demonstrated, which can effectively suppress magnetic fields, their gradients and transients \cite{Kornack2005}. SERF magnetometers and comagnetometers are currently among the most sensitive atomic magnetometers. 

\begin{figure}[bth]
\centering
\includegraphics[height=1.8in]{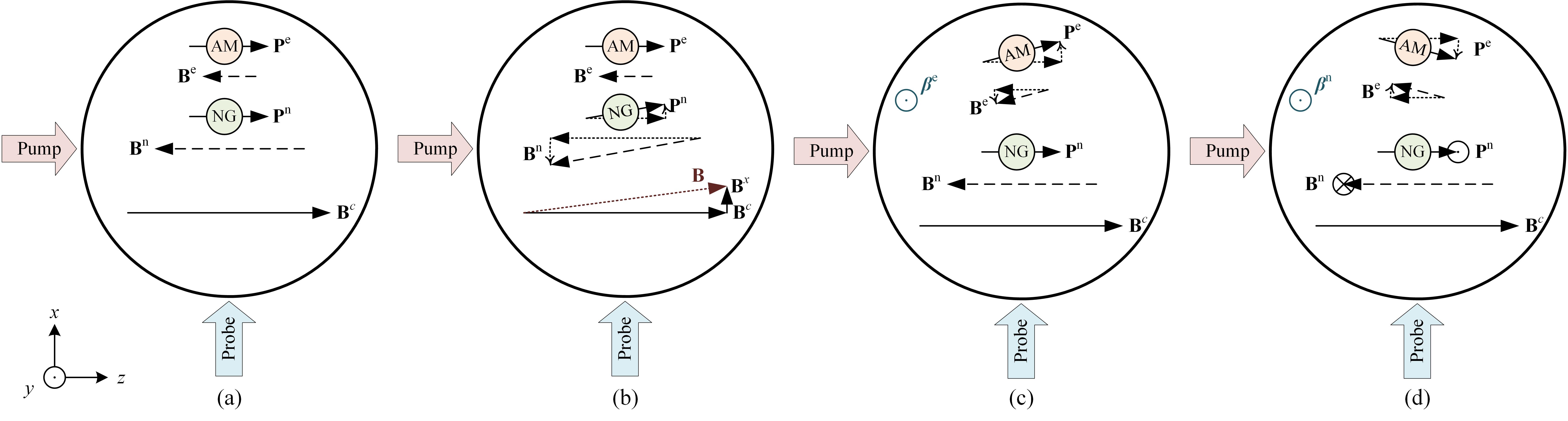}
\caption{Basic operation of self-compensating comagnetometer. (a) AM electron spins and NG nuclear spins are polarized by optical pumping along $\hat{z}$. (b) The self-compensating of the magnetic field along $\hat{x}$ by the magnetization of NG nuclear spins. (c) The spin polarization response to an anomalous field coupled only to AM electron spins. (d) The spin polarization response to an anomalous field coupled only to NG nuclear spins.
}
\label{fig:SERF}
\end{figure}

The operational principle of the SC comagnetometer is shown in Figure\,\ref{fig:SERF}. The AM electron spins are polarized along the $\hat{z}$ by a pump light, which is usually the D1 line light of AM. NG nuclear spins are polarized through spin-exchange collisions with AM electron spins as shown in Figure\,\ref{fig:SERF}(a). These two ensembles are coupled by the spin-exchange interactions between them, which can be represented by an effective magnetic field  
\begin{equation}
    \textbf{B}^{\rm{e/n}}=\lambda M^{\rm{{e/n}}}_0 \textbf{P}^{\rm{e/n}}\,,
\end{equation}
where $\lambda=8\pi\kappa_0/3$ \cite{Schaefer1989} in a uniformly polarized spherical cell and the enhancement factor $\kappa_0$ varies from 5 to 600 for different alkali-metal-noble-gas pairs \cite{Walker1989}. $M^{\rm{e/n}}_0$ represents the magnetizations of electron or nuclear spins corresponding to full spin polarizations. $\textbf{P}^{\rm{e/n}}$ denotes the spin polarization of AM electron spins or NG nuclear spins. Applying a magnetic field $B_c$ along $\hat{z}$, with the unique value of $B_c=-B^{\rm{e}}_z-B^{\rm{n}}_z$, makes the system work at the self-compensating point \cite{Kornack2005}. In this configuration, NG nuclear polarization can track small variations of magnetic fields, while keeping AM undisturbed as shown in Figure\,\ref{fig:SERF}(b). The self-compensating feature leaves AM electron spins and NG nuclear spins sensitive to anomalous fields $\boldsymbol{\beta}^{\rm{e}}$ and ${\boldsymbol{\beta}}^{\rm{n}}$ \cite{Kornack2002} . When an anomalous field ${\boldsymbol{\beta}}^{\rm{e}}$ coupling only to the AM electrons applies, AM electron spins precess with it and can be detected by a probe light based on optical rotation  as shown in Figure\,\ref{fig:SERF}(c). An anomalous field ${\boldsymbol{\beta}}^{\rm{n}}$ that couples only to the NG nuclear spins will cause the precession of NG nuclear spins, leading to the change of $\textbf{B}^{\rm{n}}$. The AM electron spins precess under this field, which can be seen in Figure \ref{fig:SERF}(d). It is remarkable that AM spins rotate in different directions under these two anomalous fields (coupling to the electrons and NG nuclei, respectively).

The Hamiltonians of AM and NG are 
\begin{equation}\label{eqn-31}
\left\{ \begin{array}{l}
\hat{H}^{\rm{e}} =A_g\textbf{I}\cdot\textbf{S}+g_e \mu_B\textbf{S}\cdot\textbf{B}-\frac{\mu_I}{I}\textbf{I}\cdot\textbf{B}+g_e\mu_B\textbf{S}\cdot{\boldsymbol{\beta}}^{\rm{e}}\,,\\
\hat{H}^{\rm{n}} =-\frac{\mu_K}{K}\textbf{K}\cdot\left(\textbf{B}+{\boldsymbol{\beta}}^{\rm{n}}\right)\,,
\end{array}\right.
\end{equation}
where $\textbf{I}$ and $\textbf{S}$ are the spin operators of AM nucleus and electron, respectively. $\textbf{B}$ is the magnetic field, $A_g$ is the ground state hyperfine coupling constant, $g_e$ is the electron spin g-factor, $\mu_B$ is the Bohr magneton, $\mu_I$ is the nuclear dipole moment of AM, and $\mu_K$ is the nuclear dipole moment of NG.
Considering these interactions, the evolution of the hybrid spin ensembles can be described by the Bloch equations
\begin{equation}\label{eqn-32}
\left\{ \begin{array}{l}
\frac{\partial{\bf{P}}^{\rm{e}}}{\partial t}= \frac{{{\gamma _{\rm{e}}}}}{Q}\left({\bf{B}}  + \lambda M_0^{\rm{n}}{{\bf{P}}^{\rm{n}}} + {{\boldsymbol{\beta}}^{\rm{e}}}\right) \times {{\bf{P}}^{\rm{e}}} + \frac{{{R_p}{{\bf{S}}_{\bf{p}}} + {R_m}{{\bf{S}}_{\bf{m}}} + R_{\rm{se}}^{\rm{ne}}{{\bf{P}}^{\bf{n}}}}}{Q} - \frac{{\{ R_2^{\rm{e}},R_2^{\rm{e}},R_1^{\rm{e}}\} }}{Q}{{\bf{P}}^{\rm{e}}}\,,\\
\frac{\partial{\bf{P}}^{\rm{n}}}{\partial t} = {\gamma _{\rm{n}}}\left({\bf{B}} + \lambda M_0^{\rm{e}}{{\bf{P}}^{\rm{e}}}+{{\boldsymbol{\beta}}^{\rm{n}}}\right) + R_{\rm{se}}^{\rm{en}}{{\bf{P}}^{\rm{e}}} - \{ R_2^{\rm{n}},R_2^{\rm{n}},R_1^{\rm{n}}\} {{\bf{P}}^{\rm{n}}}\,,
\end{array} \right.
\end{equation}
where ${\bf{P}}^{\rm{e}}$ and ${\bf{P}}^{\rm{n}}$ are the polarizations of AM electron spin and NG nuclear spins. $\gamma_e$ is the gyromagnetic ratio of AM electron spin. $Q$ is the slowing down factor. $R_p$ and $R_m$  are the mean pumping rates of unpolarized atoms of the ground state by the pump light and the probe light. ${{\bf{S}}_{\bf{p}}}$ and ${{\bf{S}}_{\bf{m}}}$ are the photon spin of pump light and probe light. $R_{\rm{se}}^{\rm{ne}}$ and $R_{\rm{se}}^{\rm{en}}$ are the spin-exchange rates experienced by AM and NG, repectively. $R_1^{\rm{e/n}}$ and $R_2^{\rm{e/n}}$ are the longitudinal and transverse relaxation rates of AM electron or NG nuclear spins.

The finally attained signal is
\begin{equation}\label{eqn-33}
P_x^{\rm{e}} = \frac{{{\gamma _{\rm{e}}}P_z^{\rm{e}}}}{{R_2^{\rm{e}}}}\left\{ {\beta_y^{\rm{e}} - \left({1+\frac{\delta B_z}{\lambda M_0^{\rm{n}} P_z^{\rm{n}}}}\right) \beta_y^{\rm{n}}-\frac{\delta B_z}{\lambda M_0^{\rm{n}} P_z^{\rm{n}}} B_y+\frac{\gamma_{\rm{e}}}{R_2^{\rm{e}}}{\delta B_z}\left[-\beta_x^{\rm{e}}+\left({1+\frac{\delta B_z}{\lambda M_0^{\rm{n}} P_z^{\rm{n}}}}\right) \beta_x^{\rm{n}} + \frac{\delta B_z}{\lambda M_0^{\rm{n}} P_z^{\rm{n}}} B_x \right] } \right\}\,.
\end{equation}

When the coupled spin ensembles work at the self-compensating point mentioned above, that means $\delta B_z=B_z - B_c =0 $, the AM electron spin is protected from magnetic field disturbance. 

\begin{figure}[bth]
\centering
\includegraphics[height=4in]{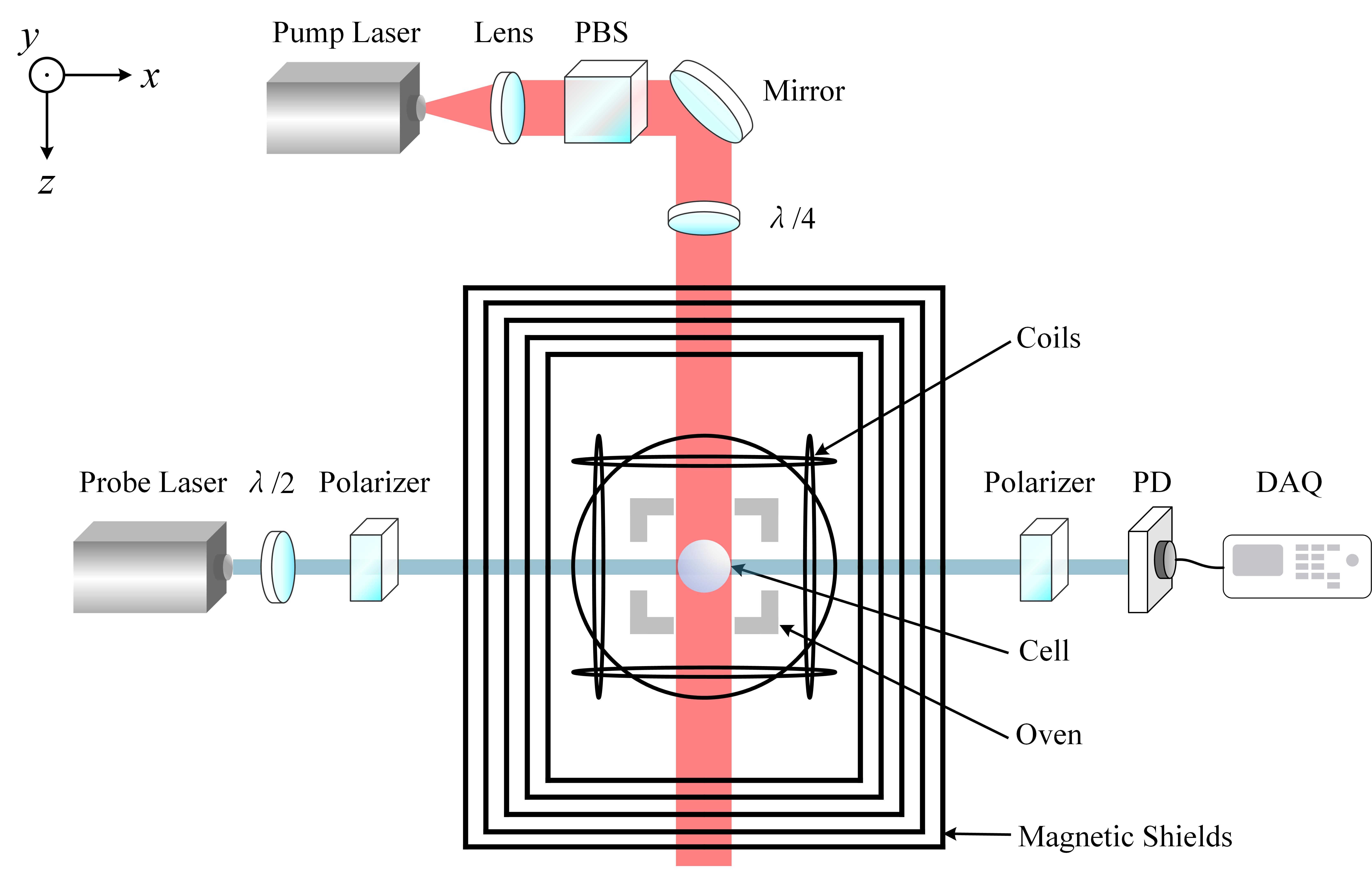}
\caption{Typical experimental setup. A pump light along $\hat{z}$ and a probe light along $\hat{x}$ are used to polarize and probe the alkali-atom spins. The vapor cell is heated with an electric heater. Magnetic shields are used to shield the cell from external magnetic fields. Three-axes magnetic coils are used to compensate residual magnetic fields and control the spin ensembles. PBS: polarization beam splitter, PD: photodetector, DAQ: Data Acquisition.}
\label{fig:Setup}
\end{figure}

A typical experimental setup is shown in Figure \ref{fig:Setup}. In the center of SC comagnetometer is a vapor glass cell, which contains the spatially overlapped AM atoms and NG atoms, as well as a small amount ($\sim$50\,torr) of $\textrm{N}_2$. AM spins are used to pump NG spins by spin-exchange optical pumping (SEOP) and acts as an in-situ magnetometer scoping the precession of noble gas which is sensitive to the exotic interactions related to ALPs. $\textrm{N}_2$ molecules quench the AM excitation, thus mitigating the deleterious effects of radiation trapping on AM polarization.
High AM atomic density and low magnetic field are two critical factors for the spin-exchange-relaxation-free regime characterized by the spin-exchange rate exceeding the Larmor-precession rate.  The atomic cell is heated to around 200\,$^\circ$C with an ac electric heater to ensure adequate alkali metal number density and avoid low-frequency  magnetic noise from the heater.  Magnetic shields are used to shield the Earth's (or laboratory) magnetic field. Typically, a multi-layer $\mu$-metal magnetic shield with a Ferrite inner layer is used to provide both high magnetic shielding factor and low magnetic noise from the shield itself. The typical residual magnetic field inside the magnetic shields is less than 2\,nT in three directions after degaussing, which are further compensated by magnetic coils. Triaxial coils are used to guarantee uniform magnetic fields and gradient magnetic fields to manipulate the spin ensembles.

A circularly polarized resonant pump light polarizes the AM electrons along $\hat{z}$.  A linearly polarized far-detuned probe light propagating along $\hat{x}$ is used to measure the transverse AM spin polarization through optical rotation. Usually, the probe light is detected with a balanced differential polarimeter to suppress common-mode noise or is modulated to high-frequency and demodulated with a lock-in amplifier to reduce low-frequency noise. Water cooling and thermal insulation are used to  prevent heating of the magnetic shield by the heat from the electric heater. The vibration isolator and vacuum chamber protect the system from vibration and air convection disturbance, respectively. Moreover, to reduce the spin-polarization gradients of AM spins due to strong light absorption along the light propagation direction and improve the hyperpolarization of NG nuclear spins, a more sophisticated hybrid SEOP is applied. Two kinds of AM atoms with different number densities are used, for example, K and Rb. The optically thin K vapor is directly spin-polarized with potassium D1 light to avoid strong light absorption. The optically thick Rb vapor is spin-polarized by spin-exchange collisions with the K atoms to achieve a uniform polarization. NG nuclear spins are further polarized by spin-exchange collisions with high-density Rb atoms to achieve a high NG polarization.

We now briefly describe various modes of operation of this basic setup. A K-$^3$He SC comagnetometer was recently used to search for ALPs \cite{Lee2023}. The result provides a new limit for  neutron-spin coupling of $2.4\times 10^{-10}\,\textrm{GeV}^{-1} $ (with a median of 95$\%$)  for most axion masses from 0.4 to 4\,$\textrm{feV}$, improving the previous laboratory bounds by about 5 orders of magnitude in that mass range. In 2022, a K-Rb-$^{21}$Ne comagnetometer and a tungsten ring featuring a high nucleon density are used to search for possible exotic spin-dependent force, specifically spin-and-velocity-dependent forces, which could be mediated by Z' particle \cite{Wei2022}. The coupling constant limit derrived from this spin-velocity-dependent force represents more than one order of magnitude tighter than that of astronomical and cosmological limits.  A K-$^3$He SC comagnetometer  with a sensitivity of $0.75\,\textrm{fT/Hz}^{1/2}$ at $0.18\,\textrm{Hz}$ is used to measure interactions with a separate optically pumped $^3$He nuclear spin source \cite{Vasilakis2009,Vasilakis2011}. The result limits the anomalous spin-spin interaction between neutrons to be smaller than $2.5\times10^{-8}$ of their magnetic interaction or under $2\times10^{-3}$ of their gravitational interaction at a length scale of $50\,\textrm{cm}$. The limit on the product of the axion pseudoscalar and scalar coupling to neutrons and nucleons was improved by an order of magnitude by using a K-$^3$He comagnetometer and two 250\,kg Pb source masses \cite{Lee2018,Lee2019}. An Rb-$^{21}$Ne comagnetometer was developed for long-range spin-dependent interactions with a rotatable SmCo$_5$ electron-spin source. The result improved the product of the pseudoscalar electron and neutron couplings and of their axial couplings by two orders of magnitude \cite{Almasi2020}.

The SC magnetometers have also been applied to test the Lorentz and CPT symmetries (not necessarily related to DM). In 2011, a K-$^3$He comagnetometer achieved an energy resolution of $10^{-34}\,\textrm{GeV}$  which improved the previous anomalous spin-dependent force of $0.05 \pm 0.56\,\textrm{aT}$ limit by a factor of 500 with a reduced $\chi^2$ of 0.87~\cite{Brown2011D}. According to the theoretical sensitivity, $^{21}$Ne could improve the energy resolution by one order of magnitude compared with $^3$He. And with nuclear spin $K$=3/2, $^{21}$Ne can be used to test local Lorentz invariance \cite{Chupp1989,Ghosh2010}. In order to suppress the Earth signal, the Cs-Rb-$^{21}$Ne comagnetometer was moved to the South pole in 2013 \cite{Smiciklas2013}. The uncertainty of one-day data was approximately equal to the Lorentz Violation limit at that time. The zero-frequency responses of pseudomagnetic signals (e.g., ALP fields, rotations) are not always the same as those in other frequency ranges. In 2023, M.\,Padniuk $\it{et\,al}$ utilized step perturbations of the transverse magnetic field to reliably predict the frequency responses to any spin perturbations. The calibration procedure is necessary for explorations using SC comagnetometers \cite{padniuk2023universal}.


\subsection{Clock-comparison comagnetometer:}
The Nobel Prize in Physics 1943 was awarded to Otto Stern ``for his contribution to the development of the molecular ray method and his discovery of the magnetic moment of the proton''. Isidor Isaac Rabi won the Nobel Prize in Physics in 1944 ``for his resonance method for recording the magnetic properties of atomic nuclei'' in 1938 \cite{Rabi1938}. Later, Felix Bloch \cite{Bloch1946} and Edward Purcell \cite{Purcell1946} independently discovered the nuclear magnetic resonance (NMR) phenomenon in 1946; they shared the Nobel Prize in Physics in 1952 ``for their development of new methods for nuclear magnetic precision measurements and discoveries in connection therewith''. The equations proposed by Bloch describing the variation of the polarization vector laid a theoretical foundation for the development of NMR technology. Since then, the Bloch equations have been used as a tool in biochemistry, medical imaging, the study of atomic interactions, and so on. 

\begin{figure}[bth]
\centering
\includegraphics[height=2.5in]{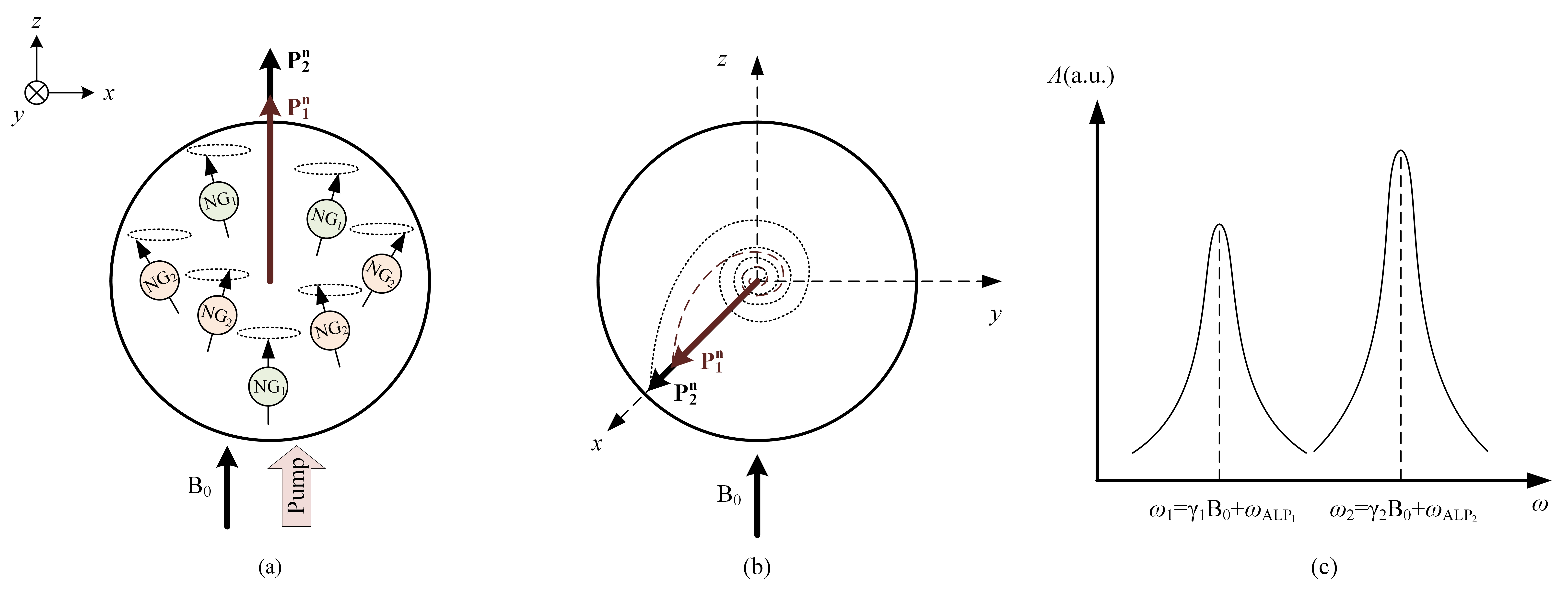}
\caption{Basic operation of clock-comparison comagnetometer. (a) Two kinds of NG nuclear spins are polarized by SEOP. (b) FID/precession around ${{\bf{B}}_0}$ after a transverse $\pi/2$ pulse along $\hat{y}$ (or other pulses that can rotate nuclear polarizations away from $\hat{z}$). (c) The spectrogram of the freely precessing nuclear spins. }
\label{fig:Clock}
\end{figure}

Magnetic noise and fluctuations have long been the primary limitations in spin-based precision measurement. The precession of noble-gas spins varies with the magnetic field at the Larmor frequency which depends on the gyromagnetic ratio. To reduce the effect of magnetic fluctuations, a clock-comparison (CC) comagnetometer was proposed \cite{Simpson1963,karwacki1980}, which uses the spin precession of two distinct NG species or different isotopes. The two noble gas spin species are polarized simultaneously by optical pumping either before they are used in the experiment or via in-situ SEOP, as shown in Figure\,\ref{fig:Clock}(a). After applying a transverse pulse field, nuclear spins undergo free induction decay (FID), as shown in Figure\,\ref{fig:Clock}(b).  An alkali-metal or a superconducting quantum interference device (SQUID) detects a nuclear spin precession signal. Two hybrid noble gases feel the same magnetic field ${\bf{B}_0}$ and anomalous interaction which has no relation to the gyromagnetic ratio, as shown in Figure\,\ref{fig:Clock}(c). Magnetic field fluctuation can be eliminated by measuring these two precession frequencies at the same time. Thus, the exotic field contribution can be extracted.

For example, under the  monopole-dipole interaction between a massive particle and a spin, which is mediated by axions, as shown in Eq.\,(\ref{eqn-23}), the frequency shift of $i$-th NG nuclear spin is \cite{Feng2022} 
\begin{equation}
    \hbar \omega_{\rm{ALP_i}}=\hbar\gamma_i{{\bf{B}}_ {i}}=\beta\left(r\right)(\hat{\bf{r}}\cdot\hat{\bf{z}})\frac{{\left\langle {\boldsymbol{\sigma}}_{\rm{n}}  \right\rangle }_i}{{\left\langle \bf{K}  \right\rangle }_i}\,,
\end{equation}
where the parameter definitions are the same as those in Eq.\,(\ref{eqn-23}).

The Bloch equations of these two NG species are 
\begin{equation}\label{eqn-clock1}
\left\{ \begin{array}{l}
\frac{\partial{\bf{P}}^{\rm{n}}_{\rm{1}}}{\partial t} = {\gamma _1}\left({\bf{B}} +{\bf{B}_1}\right) + R_{\rm{se}}^{\rm{en_1}}{{\bf{P}}^{\rm{e}}} - \{ R_2^{\rm{n_1}},R_2^{\rm{n_1}},R_1^{\rm{n_1}}\} {{\bf{P}}^{\rm{n}}_{\rm{1}}}\,,\\
\frac{\partial{\bf{P}}^{\rm{n}}_{\rm{2}}}{\partial t} = {\gamma _2}\left({\bf{B}} +{\bf{B}_2}\right) + R_{\rm{se}}^{\rm{en_2}}{{\bf{P}}^{\bf{e}}} - \{ R_2^{\rm{n_2}},R_2^{\rm{n_2}},R_1^{\rm{n_2}}\} {{\bf{P}}^{\rm{n}}_{\rm{2}}}\,,
\end{array} \right.
\end{equation}
where ${\bf{P}}^{\rm{n}}_{\rm{1}}$ and ${\bf{P}}^{\rm{n}}_{\rm{2}}$  are the polarizations of two NG nuclear spin ensembles,respectively. The indexes 1 and 2 refer to the two NG species.

By measuring the Larmor frequencies of two NG ensembles, the  desired information about  exotic interactions can be extracted according to \cite{Gemmel2010PRD}
\begin{equation}
    \Delta\omega_{\rm{ALP}}=\omega_1-\frac{\gamma_1}{\gamma_2}\omega_2=\gamma_1 {\rm{B}}_0+\omega_{\rm{ALP_1}}-\frac{\gamma_1}{\gamma_2}\left(\gamma_2 {\rm{B}}_0+\omega_{\rm{ALP_2}}\right)=\omega_{\rm{ALP_1}}-\frac{\gamma_1}{\gamma_2}\omega_{\rm{ALP_2}}\,,
\end{equation}
where the fluctuations and drifts of the magnetic field cancel out (the comagnetometer advantage). 

Nuclear spin precession in CC comagnetometers can be detected by in-situ AM electron spins. In 2013, $^{129}$Xe and $^{131}$Xe dual-species NMR frequency shifts were used to explore the long-range interactions mediated by ALPs. The Rb atoms polarized the Xe nuclei through spin-exchange collisions, and also served as a magnetometer to detect the precession of two xenon isotopes. The frequency shifts were induced by a nonmagnetic zirconia rod. Upper bounds $|g_{\rm{p}}^{\rm{n}}g_{\rm{s}}|\leq10^{-18}$ were set on the product of the coupling constants for a force range of 0.1\,mm and $|g_{\rm{p}}^{\rm{n}}g_{\rm{s}}|\leq2\times10^{-24}$ for that of 1\,cm \cite{Bulatowicz2013}. With a similar CC comagnetometer interacting with a nonmagnetic bismuth germanate (BGO) crystal, the upper bound on $|g_{\rm{p}}^{\rm{n}}g_{\rm{s}}|$ was further improved in 2022 \cite{Feng2022}. In 2023, S.-B.\,Zhang $\it{et\,al.}$ used a $^{129}$Xe-$^{131}$Xe-Rb comagnetometer to explore the ``monopole-dipole'' interaction between dual-species xenon and the Earth. The ratios between nuclear spin-precession frequencies of $^{129}$Xe and $^{131}$Xe were explored when flipping the relative direction between the sensor and Earth's gravitational field. The result expressed as a limit on $|g_{\rm{p}}^{\rm{n}}g_{\rm{s}}|$ reaches  $3.7\times10^{-36}$ ($95\%$ CL) for a force range greater than $1\times10^{8}\textrm{m}$ \cite{Zhang2023}.

SQUIDs are also sensitive sensors to detect the precession of colocated NG ensembles for ALP exploration \cite{Gemmel2010}. ALPs mediate a parity and time-reversal symmetry-violating force which enables exploring ALPs through the monopole-dipole coupling. In 2010, C.\,Gemmel $\it{et\,al}$ searched for sidereal variations based on colocated $^3$He and $^{129}$Xe ensembles while the Earth and hence the laboratory reference frame rotates with respect to a relic background field. The $^3$He and $^{129}$Xe nuclear spins were polarized outside the shielding through optical pumping. The equatorial component of the background field interacting with the spin of the bound neutron was found to be $\tilde{b}_\bot^{\rm{n}}<3.7\,\times10^{-32}\,\textrm{GeV}$ ($95\%$ CL) \cite{Gemmel2010PRD}. The result was improved by a factor of 30 in 2014 due to the higher xenon polarization, the use of four independent gradiometers and the larger size of the cell for longer longitudinal wall relaxation time \cite{Allmendinger2014}. 
In 2013, K.\,Tullney $\it{et\,al.}$ used a cylindrical BGO crystal for its low magnetism and high mass density. The nuclear spins were also polarized through optical pumping. Their result improved the previous constraints on the exotic force by four orders of magnitudes for the mediator mass heavier than 20\,$\mu\textrm{eV}$. 




\subsection{Nuclear-magnetic-resonance magnetometers:}
Ultralight bosonic particles can manifest as classical fields oscillating at the Compton frequency coupling to nuclear and electron spins and affecting the spin precession as shown in Eq.\,\ref{eqn-25} \cite{Budker2014}.  When the Larmor frequency of nuclear spins matches the oscillating field, hyperpolarized long-lived nuclear spins can greatly amplify the oscillating signal. Thus, nuclear magnetic resonance (NMR) is one of the effective ways to detect such interactions \cite{jackson_kimball_search_2023}. The time-dependent magnetization can be measured by an alkali-metal magnetometer, a pick-up coil, or a SQUID magnetometer. The NMR detected by an in-situ atom magnetometer greatly enhances the effective field generated by nuclear magnetization due to the Fermi-contact enhancement factor \cite{Schaefer1989,Jiang2021}.

The experimental setup is similar to that of the SC comagnetometer. A vapor cell made of pyrex glass or other alkali corrosion-resistance glass contains AM atoms, NG atoms and N$_2$ as buffer gas. A twisted wire carrying AC current is used to heat the vapor cell. $\mu$-metal shield is applied to shield the external magnetic field. Two sets of three pairs of orthogonal coils are positioned around the vapor cell to provide bias and oscillating magnetic fields in an arbitrary direction. The configuration of the pump and probe light is similar to that of the SC comagnetometer \cite{Jiang2021}.  

The dynamics of coupling spin ensembles are similar to that described in Eq.(\ref{eqn-32}). Suppose that the =ALPs oscillating field with an amplitude of $B_a$ is along $\hat{y}$. The effective field experienced by AM atoms from NG nuclear spins is \cite{su2022review}
\begin{equation}
  \boldsymbol{B}_{\mathrm{eff}}^{\mathrm{n}}=\frac{1}{2} \lambda M_{0}^{\mathrm{n}} P_{0}^{\mathrm{n}} \gamma_{\mathrm{n}} B_{a}\left\{\frac{T_{2}^{\mathrm{n}} \cos (2 \pi \nu_a t)+2 \pi \delta \nu {T_{2}^{\mathrm{n}}}^{2} \sin (2 \pi \nu_a t)}{1+\left(\gamma_{\mathrm{n}} B_{a} / 2\right)^{2} T_{1}^{\mathrm{n}} T_{2}^{\mathrm{n}}+(2 \pi \delta \nu_a)^{2} {T_{2}^{\mathrm{n}}}^{2}} \hat{x}+\frac{T_{2}^{\mathrm{n}} \sin (2 \pi \nu_a t)-2 \pi \delta \nu {T_{2}^{\mathrm{n}}}^{2} \cos (2 \pi \nu_a t)}{1+\left(\gamma_{\mathrm{n}} B_{a} / 2\right)^{2} T_{1}^{\mathrm{n}} T_{2}^{\mathrm{n}}+(2 \pi \delta \nu_a)^{2} {T_{2}^{\mathrm{n}}}^{2}} \hat{y}\right\}\,,
\end{equation}
where $\delta \nu=\nu_a-\nu_0$ is the detuning between the Compton frequency of the ALPs oscillating field and the Larmor frequency of NG nuclear spin $\nu_0=\gamma_{\rm{n}} B^0_z/(2\pi)$. The relaxation time of NG nuclear spins is denoted by $T_2^{\rm{n}}=1/{R^2_{\rm{n}}}$ and $T_1^{\rm{n}}=1/{R^1_{\rm{n}}}$, respectively. The total magnetic field acting on the AM electron spins is 
\begin{equation}
 \boldsymbol{B}_{\mathrm{tot}}=\boldsymbol{B}_{\mathrm{eff}}^{\mathrm{n}}+B_{a} \cos (2 \pi \nu_a t) \hat{y}\,.
\end{equation}

\begin{figure}[bth]
\centering
\includegraphics[height=2.5in]{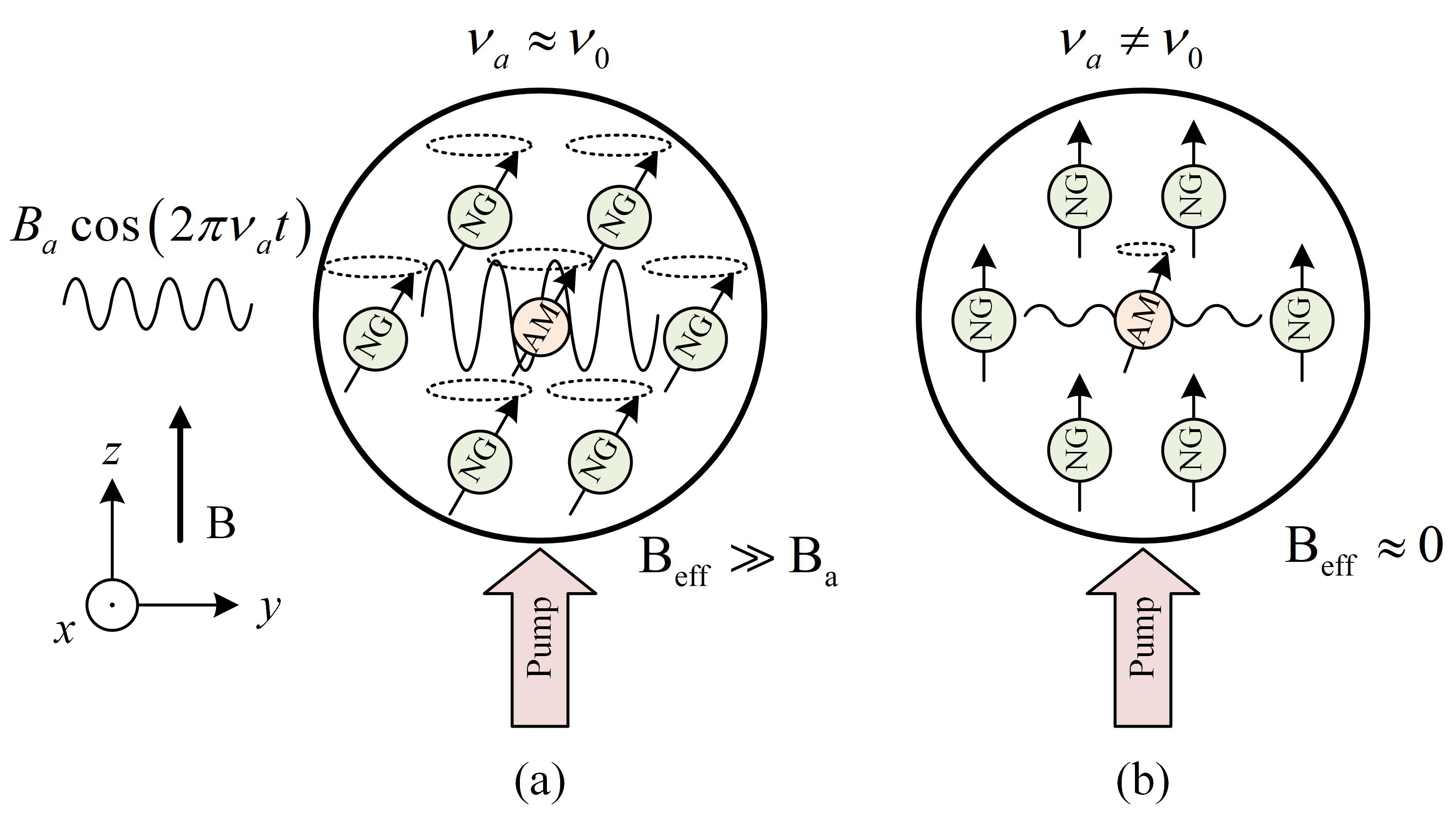}
\caption{Basic operation of  NMR magnetometer. (a) When $\nu_a$ is near the nuclear spin Larmor frequency $\nu_0$, the effective magnetic field $\boldsymbol{B}_{\mathrm{eff}}^{\mathrm{n}}$ resonantly amplifies $B_a$. Both of them affect the precession of AM electron spin. (b) When $\nu_a$ is far away from $\nu_0$, $\boldsymbol{B}_{\mathrm{eff}}^{\mathrm{n}}$ is approximately equal to 0. Only $B_a$ impacts on the electron spins. This figure is adapted from \cite{Jiang2021}}
\label{fig:SPA}
\end{figure}

It can be seen that when $\nu_a$ is in proximity to the nuclear spin Larmor frequency $\nu_0$, the effective magnetic field resonantly amplifies $B_a$. Thus, AM electron spins are influenced by $\boldsymbol{B}_{\mathrm{eff}}^{\mathrm{n}}$ which is larger than $B_a$ as shown in Figure\,\ref{fig:SPA}(a). When $\nu_a$ is far away from $\nu_0$, $\boldsymbol{B}_{\mathrm{eff}}^{\mathrm{n}}$ is approximately equal to 0. Only $B_a$ impacts the electron spins, and the influence is negligible compared to the resonant case as shown in Figure\,\ref{fig:SPA}(b). Thus, such sensor is sensitive to the ALPs oscillating field consistent with the resonant frequency.

Assume that the large bias field is along $\hat{z}$, ${\bf{B}}\approx B^0_z$ and $R^{\rm{e}}_2\approx R^{\rm{e}}_1=1/T^{\rm{e}}$. The detected signal from a probe light along $\hat{x}$ is
\begin{equation}
    P_{x}^{\mathrm{e}} \propto \frac{B_{x} B_{z}-B_{y} \left(1/\left(\gamma_{\rm{e}} T_{\rm{e}}\right)\right)}{|\boldsymbol{B}|^{2}+\left(1/\left(\gamma_{\rm{e}} T_{\rm{e}}\right)\right)^{2}} \approx \frac{B_{x} B_{z}-B_{y}\left(1/\left(\gamma_{\rm{e}} T_{\rm{e}}\right)\right)}{\left(B_{z}^{0}\right)^{2}+\left(1/\left(\gamma_{\rm{e}} T_{\rm{e}}\right)\right)^{2}}\,.
\end{equation}

In 2021, M.\,Jiang $\it{et\,al}$ searched ALPs through axion-nucleon interactions where ALPs act as an oscillating magnetic field. An NMR magnetometer with a $^{129}$Xe preamplifier and $^{87}$Rb magnetometer  realized a sensitivity of 18\,fT/Hz$^{1/2}$. By adjusting the bias field, the ALP frequencies range from 2 to 180\,Hz, in other words, the ALP masses range from 8.3 to 744.0\,feV. The constraint on coupling of ALPs to nucleons was set to $2.9\times 10^{-9} \,\rm{GeV}^{-1}$ at 67.5\,feV ($95\%$ CL) \cite{Jiang2021}. The gradient of the ALP field oscillates in a narrow band of frequencies, the strength of which is related to the energy density and velocity distribution of dark matter, $v_{\rm DM}$. In 2022, Itay\,M.\,Bloch $\it{et\,al.}$ used nuclei of $^{3}$He and $^{39}$K atoms which operated in the SERF regime to search for the coupling of ALPs to neutrons and protons. By scanning the magnetic field to search for the ALPs fields in the range of 0.33 to 50 kHz, ALP-neutron and ALP-proton coupling were improved two orders of magnitude for many masses in the range of $1.4\times10^{-12} \textrm{eV}/c^{2}$ to $2\times10^{-10} \textrm{eV}/c^{2}$ (from Noble And Alkali Spin Detectors for Ultralight Coherent darK matter (NASDUCK) collaboration dubbed NASDUCK-SERF )\cite{Bloch2022}.

In 2021, H.\,Su $\it{et\,al.}$ used a BGO insulator mass to generate the spin- and velocity-dependent interactions. This pseudo-magnetic field was resonantly amplified by $^{129}$Xe spins. The results set  constraints on spin-dependent interactions for forces ranging from 0.04 to 100\,m and on velocity-dependent interactions for forces ranging from 0.05 to 6\,m ($95\%$ CL) \cite{Su2021}.  In 2022, Y.\,Wang $\it{et\,al}$ used a spin source consisting of polarized $^{87}$Rb atoms to search for exotic dipole-dipole interaction. An optical chopper was used to periodically modulate the polarization of the spin source. The pseudomagnetic field mediated by axions was resonantly amplified by hyperpolarized $^{129}$Xe and detected by $^{87}$Rb spins. The upper bound of coupling constants $g_{\rm{p}}^{\rm{e}} g_{\rm{p}}^{\rm{n}}$ was improved for the mass range from $30 \mu\textrm{eV}$ to $1 \textrm{meV}$. At 0.1\,meV, the constraint $|g_{\rm{p}}^{\rm{e}} g_{\rm{p}}^{\rm{n}}|/4$ was set to $5.3\pm 48.5_{\rm{stat}}\pm 2.4_{\rm{syst}}$ \cite{Wang2022}.



\subsection{CASPEr:}
\begin{figure}[bth]
\centering
    \includegraphics[height=1.5in]{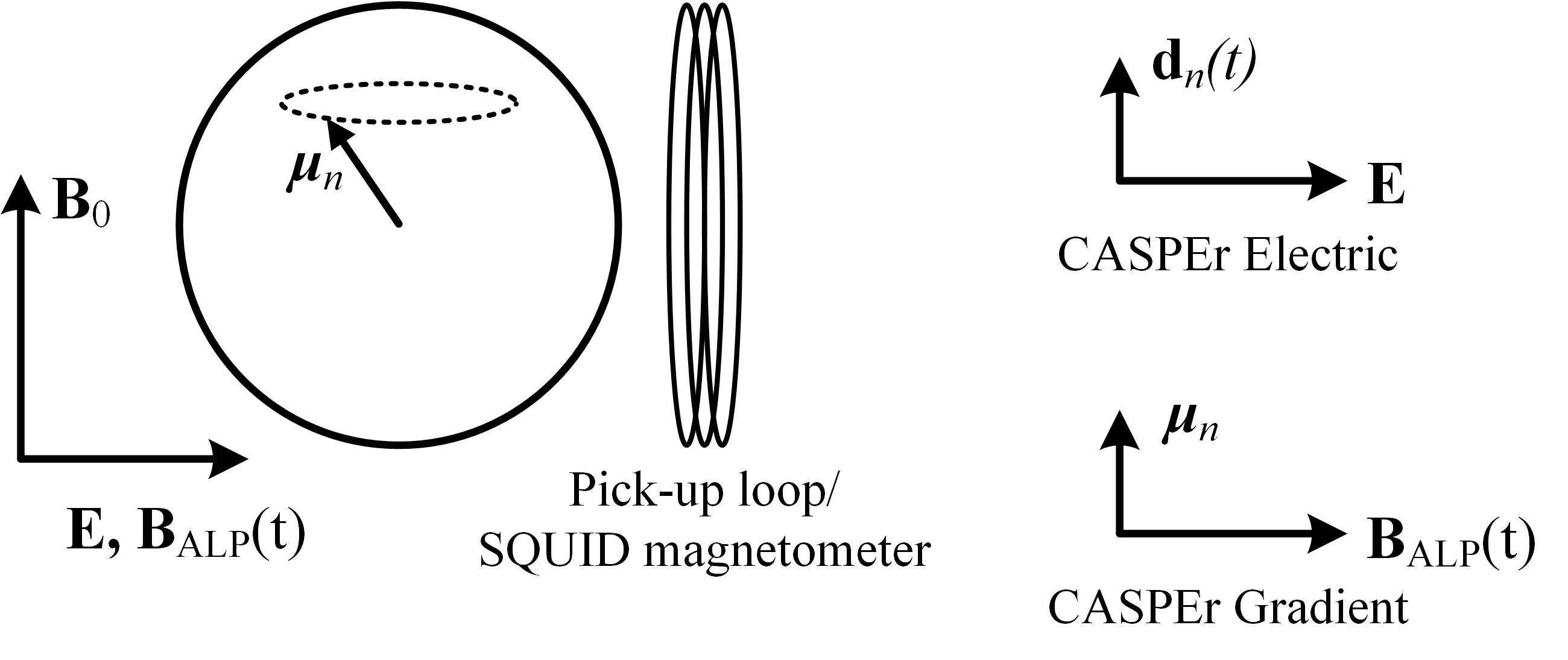}
\caption{Basic operation of the CASPEr experiment is shown on the left. When the Larmor frequency is approximately equal to the Axion field frequency, axion-induced torque tilts the nuclear spins away from the initial orientation. This figure is adapted from \cite{jackson_kimball_search_2023}}
\label{fig:Casper}
\end{figure}

The ALP field $a\left( \boldsymbol{r},t\right)$, volunteered for DM, can couple to standard-model spins ${{{\boldsymbol{\hat \sigma }}}_{\rm{n}}}$ through EDM interaction ${\bf{d}}_{\rm{n}}\left( t \right)$ or axion-fermion gradient interaction ${{\rm\bf{B}}}_{\rm{ALP}}\left( t \right)$ as shown in Figure\,\ref{fig:Casper} \cite{Graham2013}. 
The oscillating torques on spins are \cite{JacksonKimball2020}
\begin{equation}\label{eqn-Casper}
\left\{ \begin{array}{l}
{\boldsymbol{\tau}}_{\rm{EDM}}={\bf{d}}_{\rm{n}}\left( t \right) \times \boldsymbol{E}^*\,,\\
{\boldsymbol{\tau}}_{\rm{grad}}={\boldsymbol{\mu}}_{\rm{n}} \times {{\rm\bf{B}}}_{\rm{ALP}}\left( t \right)\,,
\end{array} \right.
\end{equation}
where $\boldsymbol{E}^*$ is the effective electric field, which depends on the atomic and nuclear structure of the spin system \cite{Aybas2021}. $\boldsymbol{\mu}_{\rm{n}} \propto \hat{\boldsymbol{\sigma}}_{\rm{n}}$ represents the nuclear magnetic moment.
The Cosmic Axion Spin Precession Experiment (CASPEr) is naturally divided into CASPEr Electric and CASPEr Gradient to explore these oscillating torques using magnetic resonance \cite{Budker2014, Wu2019, Garcon2019constraints}. By scanning the bias magnetic field, the Larmor frequency is regulated. When the Larmor frequency matches with the ALP oscillating frequency, a time-dependent magnetization of NG is generated and can be read out through a pick-up loop or with a SQUID magnetometer \cite{jackson_kimball_search_2023}. 

\subsection{Floquet NMR Magnetometer:}
Although the effective field induced by noble-gas polarization can enhance the weak external oscillating  field, the amplified signal is restricted to around the Larmor frequency of NG nuclear spins. Periodically driving the noble-gas spins  by applying an oscillating field  $B_{\rm{ac}}\cos(2\pi\nu_{\rm{ac}} t)\hat{z}$ parallel to the bias field, namely, a Floquet field, can enable measurements at multiple different frequency regimes \cite{Jiang2022}. 

The effective field under the condition of the Floquet field is \cite{su2022review}
\begin{equation}
    \boldsymbol{B}_{\mathrm{f}, \mathrm{eff}}^{\mathrm{n}}=\lambda M_{0}^{\mathrm{n}} B_{a}\left\{\sum_{l=-\infty}^{+\infty} \sum_{k=-\infty}^{+\infty} B_{k, l}(u, \nu) \cos \left[2 \pi\left(\nu+l \nu_{\mathrm{ac}}\right) t\right]+A_{k, l}(u, \nu) \sin \left[2 \pi\left(\nu+l \nu_{\mathrm{ac}}\right) t\right]\right\}\,,
\end{equation}
where the coefficients are defined as 
\begin{equation}
\left\{ \begin{array}{l}
A_{k, l}(u, \nu)=\frac{\gamma_{n} P_{0}^{\mathrm{n}} T_{2 \mathrm{n}} J_{k+l}(u) J_{k}(u)}{2} \frac{1}{1+\left[2 \pi\left(k v_{\mathrm{ac}}-\delta \nu\right) T_{2 \mathrm{n}}\right]^{2}}\,,\\
B_{k, l}(u, \nu)=\frac{\gamma_{n} P_{0}^{\mathrm{n}} T_{2 \mathrm{n}} J_{k+l}(u) J_{k}(u)}{2} \frac{2 \pi\left(k v_{\mathrm{ac}}-\delta \nu\right) T_{2 \mathrm{n}}}{1+\left[2 \pi\left(k v_{\mathrm{ac}}-\delta \nu\right) T_{2 \mathrm{n}}\right]^{2}}\,,
\end{array} \right.
\end{equation}
which contains the modulation index $u=(\gamma_{\rm{n}} B_{\rm{ac}})/{\nu_{\rm{ac}}}$ and the first kind of Bessel function $J_k$. The Floquet system can reach multiple resonances at the frequencies that satisfy $\delta \nu = k\nu_{\rm{ac}}$.

The experimental setup of the Floquet NMR magnetometer is similar to that of the NMR magnetometer, with the main distinction being the addition of a longitudinal periodically driving field. In 2022, M. Jiang $\it{et\,al}$ applied an oscillating field to periodically drive the $^{129}$Xe spins and achieved the magnetic field amplification at a series of comelike frequencies correspongding to transitions between Floquet states. The multiresonance was in-situ read out by an $^{87}$Rb magnetometer. The magnetic sensitivity of the Floquet-amplification was approximately 20\,fT/Hz$^{1/2}$, 25\,fT/Hz$^{1/2}$, and 18\,fT/Hz$^{1/2}$ at $\nu_0 -\nu_{\rm{ac}}$, $\nu_0$, $\nu_0 +\nu_{\rm{ac}}$, respectively \cite{Jiang2022}.

Focusing on the significant gap between electron and nuclear gyromagnetic ratio, the different resonance frequencies between these two species at high bias field limit  sensitivity improvement. Applying a Floquet field  introduces multiple resonance frequencies to the alkali electron spins, which enhances the response of the ALP oscillating field and improves the sensitivity. 
NASDUCK collaboration searched for the oscillating magnetic-like field coupled of ALPs to spins using polarized $^{129}$Xe ensembles and in-situ $^{85}$Rb Floquet magnetometer. Based on the measurements over 5 months, the mass range of ALP-neutron interactions of $4\times10^{-15}$ to $ 4\times10^{-12}\,\textrm{eV}/c^2$ was further limited, and improved the bounds by up to 3 orders of magnitude for masses above $4\times10^{-13}\,\textrm{eV}/c^2$ \cite{Bloch2022SA}.


\subsection{Maser:}
\begin{figure}[bth]
\centering
\includegraphics[height=2.3in]{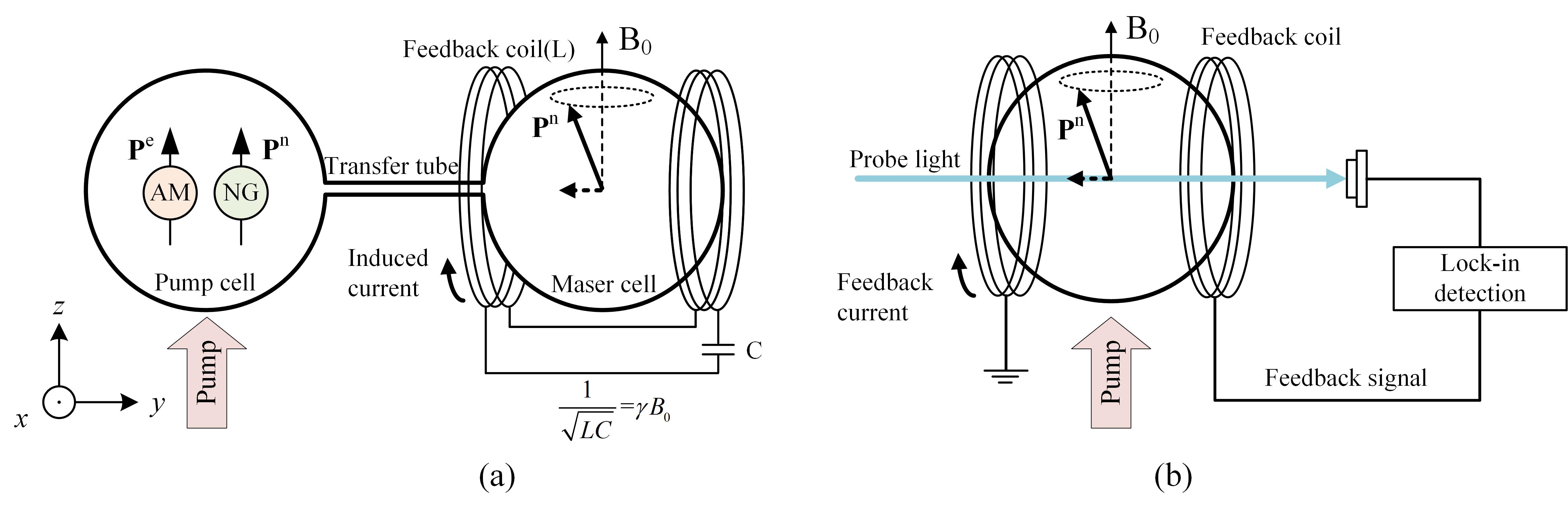}
\caption{Basic operation of spin Maser. (a) A conventional nuclear spin maser. A diffusion tube connects the pump cell and the maser cell. Persistent oscillating fields are sustained through the continuous pumping and the pick-up coils with resonant circuits. (b) An artificial feedback spin maser. The feedback field is related to the transverse component of nuclear polarization. This figure is adapted from  \cite{Yoshimi2002}}
\label{fig:maser}
\end{figure}

The first NG Zeeman maser was the $^{3}$He maser in 1964 \cite{Robinson1964}. Sustained maser oscillation implies an infinite lifetime of nuclear spins, which is beneficial for sensitivity \cite{Richard1988}. The conventional type of nuclear spin maser is shown in Figure\,\ref{fig:maser}(a). The pump cell and the maser cell are connected by the diffusion tube which guarantees different conditions for these two cells simultaneously. A relatively high temperature is required in the pumping cell for spin-exchange optical pumping, while the maser cell needs to maintain the long coherence time of the nuclear spins. The pick-up coils with resonant circuits can generate persistent oscillating fields under continuous pumping. Magnetic field adjustment and ALP information can be obtained by lock-in amplifier. Two noble-gas-species configuration is used to remove the magnetic field interference. 

The maser is a powerful tool for the measurement of EDM  as well as for the exploration of DM. In 2001, M. A. Rosenberry $\it{et\,al.}$ utilized the stable oscillations of spin-exchange pumped maser of $^{129}$Xe and $^{3}$He to  achieve a sensitivity of $d(^{129}{\rm{Xe}})=(+0.7\pm 3.3)\times 10^{-27}\,e \textrm{cm}$, which is 4-fold higher than the result in 1984 \cite{Rosenberry2001}. 
In 2008, Alexander G. Glenday $\it{et\,al.}$ used a $^{3}$He/$^{129}$Xe maser to search for new couplings between neutron spins. Through monitoring the nuclear Zeeman frequencies of the $^{3}$He/$^{129}$Xe maser while modulating a polarized $^{3}$He spin source, the coupling strength $|g_{\rm{p}}^{\rm{n}} g_{\rm{p}}^{\rm{n}}|/(4\pi)$ was limited under $3\times10^{-7}$ for interaction range more than about 41 cm in 2008 \cite{Glenday2008}.

Another type of nuclear spin maser operates through an artificial feedback mechanism, enabling self-sustained precession of spins at low density as shown in Figure\,\ref{fig:maser}(b). The transverse component of nuclear polarization is detected and feedback is controlled by a transverse magnetic field \cite{Yoshimi2002}. One specific configuration is that the transverse feedback field ${\bf{B}}_{\rm{fb}}\left( t \right) =\left( B_x\left( t \right),\,B_y\left( t \right) \right)$ is proportional to the transverse polarization of NG nuclear spins ${\bf{P}}_{\rm{T}} \equiv \left( P_x^{\rm{n}},\,P_y^{\rm{n}} \right) $ with a phase shifted by $90^\circ$, as 
\begin{equation}
\left\{\begin{array}{l}

B_x\left( t \right)=\frac{1}{\gamma_{\rm{n}}\tau_{\rm{fb}}} \frac{P_y^{\rm{n}}}{P_0} \,,\\
B_y\left( t \right)=-\frac{1}{\gamma_{\rm{n}}\tau_{\rm{fb}}} \frac{P_x^{\rm{n}}}{P_0} \,,
    \end{array}\right.
\end{equation}
where, $\tau_{\rm{fb}}$ is determined by the gain of the feedback system, and $P_0$ is the equilibrium nuclear polarization along $\hat{z}$ without feedback field. The magnetic field experienced by NG atoms is \cite{Yoshimi2002}
\begin{equation}
{\bf{B}}=B_0 \hat{z}+{\bf{B}}_{\rm{fb}}=\left(B_x\left( t \right),\,B_y\left( t \right),\,B_0\right)\,.
\end{equation}
The Bloch equations of nuclear spin maser are \cite{Sato2018}
\begin{equation}
\left\{\begin{array}{l}

\frac{\mathrm{d} P_x^{\rm{n}}}{\mathrm{d} t}=\gamma_{\rm{n}}\left[P_y^{\rm{n}}(t) B_0-P_z^{\rm{n}}(t) B_y(t)\right]-\frac{P_x(t)}{T_2^{\rm{n}}}\,, \\
\frac{\mathrm{d} P_y^{\rm{n}}}{\mathrm{d} t}=\gamma_{\rm{n}}\left[P_z^{\rm{n}}(t) B_x(t)-P_x^{\rm{n}}(t) B_0\right]-\frac{P_y(t)}{T_2^{\rm{n}}}\,,\\
\frac{\mathrm{d} P_z^{\rm{n}}}{\mathrm{d} t}=\gamma_{\rm{n}}\left[P_x^{\rm{n}}(t) B_y(t)-P_y^{\rm{n}}(t) B_x(t)\right]-\frac{P_z(t)-{\bf{P}}_0}{{T_1^{\rm{n}}}^*}\,,

    \end{array}\right.
\end{equation}
where ${T_1^{\rm{n}}}^*$ is the effective longitudinal relaxation time, including the spin exchange collision between NG and AM. Under the condition of continuous pumping and the feedback field, the NG nuclear spin polarization continuously precesses along ${\bf{B}}_{\rm{0}}$ and the ${\bf{P}}_{\rm{T}}\left( t \right)$ rotates at an angular frequency of $-\gamma_{\rm{n}} B_0$ without loss of transverse polarization. The ${\bf{P}}_{\rm{T}}$ in the stationary state is
\begin{equation}
    \begin{array}{l}
{\bf{P}}_{\mathrm{T}}=\tau_{\mathrm{fb}} \sqrt{\frac{1}{{T_1^{\rm{n}}}^*}\left(\frac{1}{\tau_{\mathrm{fb}}}-\frac{1}{T_2^{\rm{n}}}\right)} {\bf{P}}_0\,.

    \end{array}
\end{equation}


In 2012, A.\,Yoshimi $\it{et\,al.}$ reported a $^{129}$Xe nuclear spin maser that offered a better frequency width and lower operation frequencies than the conventional masers. The frequency precision of 9.3\,nHz is obtained in a measurement time of $3 \times 10^{4}\,\rm{s}$ which can be used to search for EDM \cite{Yoshimi2012}. In 2014, Y.\,Ichikawa $\it{et\,al.}$ demonstrated the EDM measurement using a $^{129}$Xe nuclear spin maser and a $^{3}$He comagnetometer. The $^{3}$He comagnetometer was used to cancel out the external magnetic field drifts. A double-cell geometry was used to deal with the drifts from contact interaction with Rb electron spins \cite{Ichikawa2014}. In 2018, T.\,Sato $\it{et\,al.}$ used colocated $^{129}$Xe and $^{131}$Xe nuclear spin masers aimed to new physics explorations. The frequency drifts from external magnetic field were reduced by two orders of magnitude. The frequency precision reached 6.2\,nHz for a $10^{4}\,\rm{s}$ average time \cite{Sato2018}.


\subsection{GNOME:}
\begin{figure}[bth]
\centering
\includegraphics[height=4in]{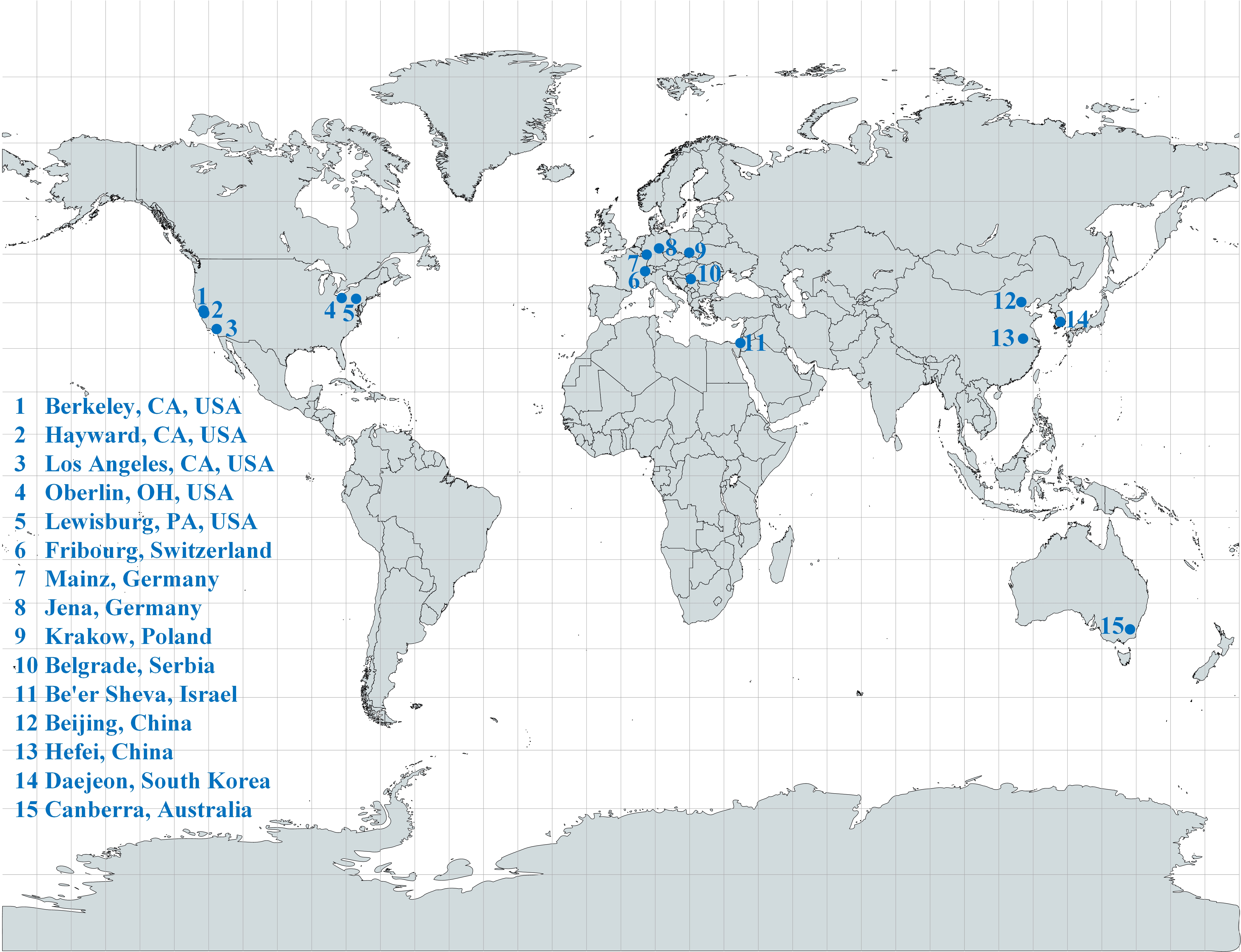}
\caption{Map and locations of GNOME magnetometers. The locations of GNOME magnetometers are marked by blue dots.  This figure is adapted from \cite{Afach2023}}
\label{fig:GNOME}
\end{figure}

The Global Network of Optical Magnetometers to search for Exotic physics (GNOME) is a network with a global distribution of sensors which is significant for the measurement of transient events attributable to exotic physics. The atom haloscopes are located in the USA, Switzerland, Germany, Poland, Serbia, Israel, China, South Korea, and Australia as shown in Figure\,\ref{fig:GNOME}. The magnetometers initially included the SERF magnetometer, rf-driven magnetometer, and AM NMOR (nonlinear magneto-optical rotation) magnetometer at first \cite{Afach2018,Afach2021}. The advanced GNOME is developing comagnetometers to reduce the magnetic disturbances  \cite{Kornack2002,Terrano2021}. Five periods of collected data  have been used to explore axion domain walls, axion stars, Q-balls, dark-matter field fluctuations, solar axion halo, Exotic Low-mass Fields (ELFs) emitted from black hole mergers and other exotic physics scenarios \cite{Afach2023}.


\subsection{Hybrid Spin Resonance:}
The bias magnetic field used to match the nuclear spin NMR frequency with the ALPs field leads to increased spin-exchange relaxation and a narrow bandwidth. Thus, the frequency scanning step is relatively small, rendering resonant searches time-consuming. A project aiming to search for light DM and other new physics (named ChangE) demonstrated a strongly-coupled Hybrid Spin-Resonance (HSR) regime of the alkali-noble-gas sensor~\cite{Wei:2023rzs}.
Hybridizing the NG and AM spins can realize an improvement in the bandwidth of NG nuclear spins \cite{Xu2022}. By adjusting the bias field to the HSR point $-B^{\rm{n}}_z$ rather than the self-compensating point of the SC comagnetometer, the Larmor frequency of AM electron spins can be slowed down while that of NG nuclear spins is accelerated. Thus, the resonant frequencies of AM and NG are moved closely with each other. Under the HSR regime, the dynamics of these two spin ensembles are strongly coupled. The bandwidth of $^{21}$Ne is improved by three orders of magnitude over the intrinsic resonance linewidth of $^{21}$Ne. Hybrid spin-exchange optical pumping was used to improve the uniformity of AM spin polarization and the hyperpolarization efficiency of NG nuclear spins. The optical-thin alkali atoms (K atoms) were directly polarized by a pumping light. The optical-thick alkali atoms (Rb atoms) are spin-polarized via spin-exchange collisons with K atoms. The NG nuclear spins are polarized by spin-exchange collisions with alkali atoms. The HSR sensor was used to search for DM. The result constrained the DM interactions with neutrons of $|g_{\rm{ann}}|\leq 3\times 10^{-10} ~{\rm GeV}^{-1}$ in the frequency range between 0.02 to 4 Hz, which surpassed the astrophysical limits. The ChangE HSR also provided the best terrestrial constraints of axion-proton coupling below 700 Hz \cite{Wei:2023rzs}.

\section{Discussion}
Although these  magnetometers and comagnetometers are all based on alkai-noble-gas hybrid spin ensembles, they exhibit quite different performances under various operation regimes. Here we summarize their main features and compare the main differences in Table.\,\Ref{tab:my_label-1}. We use $\gamma_{\rm{n}} B \hbar$ to calculate the short-term energy resolution.
\begin{table}[h]
    \renewcommand{\arraystretch}{1}
    \begin{center}
        \footnotesize

            
        \begin{tabular}{ p{1.25cm}<{\centering} p{ 3.75cm}<{\centering}  p{1.5cm}<{\centering} p{2.75cm}<{\centering} p{2.25cm}<{\centering} p{0.25cm}<{\centering} p{3.25cm}<{\RaggedRight} }

            \toprule
            Mode & Feature & Spin species   & Sensitivity (or Precision) & Energy resolution & Ref. & Applications  
            \\ \hline 
               \multirow{6}{*}{\centering  SC}  & \multirow{6}{3.75cm}{The nuclear spins of NG protect the electron spins of AM from low-frequency magnetic field fluctuations, while remaining ultrahigh sensitivity to exotic interactions.}  & \multirow{3}{*}{\centering K-$^{3}$He} & \multirow{3}{2.75cm}{\centering 0.75 fT/Hz$^{1/2}\,@$ 0.18 Hz} & \multirow{3}{*}{\centering $1.0\times 10^{-22}\, \rm{eV/Hz^{1/2}}$}  & \multirow{3}{*}{\centering \cite{Vasilakis2009}}& \multirow{6}{3.25cm}{ALP search \cite{Klinger2023}, Fifth force search \cite{Almasi2020}, CPT \cite{Kornack2005,Vasilakis2009,Vasilakis2011} and Lorentz symmetry test \cite{Smiciklas2011}*} \\
               &&&&&&\\
               &&&&&&\\
               &&\multirow{3}{*}{\centering K-Rb-$^{21}$Ne}&\multirow{3}{3cm}{\centering $3\times 10^{-8}$ rad/s/Hz$^{1/2}$\newline$@$ 0.2 to 1.0 Hz}   & \multirow{3}{*}{\centering $2.1\times 10^{-23}\, \rm{eV/Hz^{1/2}}$}&\multirow{3}{*}{\centering \cite{Wei2023}}&\\
               &&&&&&\\
               &&&&&&\\
             \hline

            \multirow{4}{1.5cm}{\centering  CC}  & \multirow{4}{3.75cm}{ Comparing the precession frequencies of colocated NG ensembles cancels the magnetic fluctuation.}  & \multirow{2}{*}{$^{3}$He-$^{129}$Xe} & \multirow{2}{3cm}{\centering 2.0$\times 10^{-28}\,e$ cm \newline(Systematic Error)}  & \multirow{2}{*}{\centering  --} & \multirow{2}{0.25cm}{\centering \cite{Sachdeva2019}} & \multirow{4}{3.25cm}{ALP search \cite{Gemmel2010}, Fifth force search \cite{Bulatowicz2013,Feng2022}, CPT and Lorentz symmetry test \cite{Allmendinger2014}, EDM \cite{Sachdeva2019}} \\
            &&&&&&\\
            & & \multirow{2}{*}{\centering $^{129}$Xe-$^{131}$Xe} & \multirow{2}{3cm}{\centering $1\times 10^{-7}$Hz/hr$^{1/2}$ $@\,{\Omega_m}/2\pi$}  & \multirow{2}{*}{\centering --}& \multirow{2}{*}{\centering \cite{Zhang2023}} \\
            &&&&&&

            \\ \hline  
            \multirow{3}{1.5cm}{\centering NMR (SAPPHIRE)} & \multirow{3}{3.75cm}{Nuclear magnetic resonance preamplifies the exotic field when it matches the Larmor frequency.}   & \multirow{3}{*}{\centering $^{87}$Rb-$^{129}$Xe}  & \multirow{3}{3cm}{\centering 18 fT/Hz$^{1/2}\,@$ $\sim$ 9 Hz} & \multirow{3}{*}{\centering $8.8\times 10^{-22}\, \rm{eV/Hz^{1/2}}$} & \multirow{3}{*}{\centering \cite{Jiang2021}} & \multirow{3}{3.25cm}{ALP search \cite{Jiang2021,Bloch2022}, Fifth force search \cite{Su2021}} \\
            &&&&&&\\

            \\ \hline 
            \multirow{4}{1.5cm}{\centering NMR (NASDUCK SERF)} & \multirow{4}{3.75cm}{The sensitivity of anomalous is improved by setting the EPR frequency near the searched ALP field.} & \multirow{4}{*}{\centering $^{37}$K-$^{3}$He} & \multirow{4}{2.75cm}{\centering \,\,\,\,\,\,\,\,\,\,$\sim$1-3 fT/Hz$^{1/2}$\newline$@$ 2-32 kHz}  & \multirow{4}{*}{\centering --} & \multirow{4}{*}{\centering \cite{Bloch2022}} & \multirow{4}{3.25cm}{ALP search, strong CP problem \cite{Bloch2022}}\\
            &&&&&&\\
            &&&&&&\\
            \\ \hline  
            
            \multirow{4}{1.5cm}{\centering NMR (ChangE)} & \multirow{4}{4cm}{The extension of measurement bandwidth can be attained by increasing the amplification factor or lowering the nonmagnetic noise.} & \multirow{4}{*}{\centering K-Rb-$^{21}$Ne}  & \multirow{4}{*}{\centering 1.3 fT/Hz$^{1/2}\,@$ 5.25Hz}  & \multirow{4}{*}{\centering $1.8\times 10^{-23}$ eV/Hz$^{1/2}$} & \multirow{4}{*}{\centering \cite{xu2023constraining}}  & \multirow{4}{3.25cm}{ALP search, CPT and Lorentz symmetry test \cite{xu2023constraining}}\\
            &&&&&&\\
            &&&&&&\\
    
            \\ \hline 

            \multirow{3}{1.5cm}{\centering NMR (CASPEr)} & \multirow{3}{3.75cm}{NMR techniques are used to explore the time-varying torque from oscillating an axion field.} 
            &\multirow{3}{*}{\centering Liquid Xe} & \multirow{3}{*}{\centering --}   & \multirow{3}{*}{\centering --} & \multirow{3}{*}{\centering \cite{Garcon_2018}} &\multirow{3}{*}{ALP search \cite{JacksonKimball2020}}\\
            &&&&&&\\
            \\ \hline 
   
            \multirow{3}{1.5cm}{\centering NMR (Floquent SAPPHIRE)} & \multirow{3}{3.75cm}{A strong Floquet field periodically driving spins results in multiple NMR amplifier.} & \multirow{3}{*}{\centering $^{87}$Rb-$^{129}$Xe}  & \multirow{3}{3cm}{\centering 20\,fT/Hz$^{1/2}$, 25\,fT/Hz$^{1/2}$, 18\,fT/Hz$^{1/2}\,@ \nu_0 -\nu_{\rm{ac}}$, $\nu_0$, $\nu_0 +\nu_{\rm{ac}}$}  & \multirow{3}{2.5cm}{\centering $8.8 \sim 12.3 \times 10^{-22}\, \rm{eV/Hz^{1/2}}$} & \multirow{3}{*}{\centering \cite{Jiang2022}} & \multirow{3}{3.25cm}{ALP search \cite{Jiang2022,Bloch2022SA}}\\
            &&&&&&\\
            \\ \hline 
            \multirow{4}{1.5cm}{\centering NMR (Floquet NASDUCK)} & \multirow{4}{3.75cm}{The strong Floquet field bridges the NMR and EPR frequency gap.} & \multirow{4}{*}{\centering $^{85}$Rb-$^{129}$Xe} & \multirow{4}{*}{\centering $\sim$45 fT/Hz$^{1/2}\,@$ 116Hz}  & \multirow{4}{*}{\centering $2.2\times 10^{-21}\, \rm{eV/Hz^{1/2}}$} & \multirow{4}{*}{\centering \cite{Bloch2022SA}} & \multirow{4}{3.25cm}{ALP search \cite{Bloch2022SA}}\\
            &&&&&&\\
            &&&&&&\\

            \\ \hline 
            
            \multirow{4}{1.5cm}{\centering Maser} & \multirow{4}{3.75cm}{With resonant circuits or feedback fields, the nuclear spins of NG can precess without losing transverse polarization.} & \multirow{4}{*}{\centering $^{129}$Xe-$^{131}$Xe}  & \multirow{4}{2.5cm}{\centering \,\,\,\,\,\,\,\,\,\,6.2$\mu\rm{Hz}$\newline (frequency precision)}  & \multirow{4}{*}{\centering --} & \multirow{4}{*}{\centering \cite{Sato2018}} & \multirow{4}{3.25cm}{ALP search, Fifth force \cite{Glenday2008}, EDM \cite{Yoshimi2012,Ichikawa2014,Sato2018}}\\
            &&&&&&\\
            &&&&&&\\

            \\ \hline 
            \multirow{4}{1.5cm}{\centering  HSR (ChangE)} & \multirow{4}{3.75cm}{Coupling nuclear spins of NG to electron spins of AM results simultaneous wide bandwidth and ultrahigh sensitivity.} & \multirow{4}{*}{\centering K-Rb-$^{21}$Ne}  & \multirow{4}{3cm}{\centering 0.78 fT/Hz$^{1/2}$\\$@$28-32 Hz}  & \multirow{4}{*}{\centering $2.0 \times 10^{-22}$ eV/Hz$^{1/2}$} &  \multirow{4}{*}{\centering \cite{wei2023dark}} & \multirow{3}{3.25cm}{ALP search, CPT and Lorentz symmetry test \cite{wei2023dark}}\\
            &&&&&&\\
            &&&&&&\\

            \\ \hline


 
        \end{tabular}
    \end{center}
    \caption{The performance comparison. SC: self-compensating comagnetometer; CC: clock-comparison comagnetometer; HSR: hybrid spin resonance; EPR: electron paramagnetic resonance; QEM: quantum-enhanced magnetometry. *A K-Rb-$^{21}$Ne self-compensating comagnetometer was developed for the test of local Lorentz invariance, the short-term energy resolution was not published~\cite{Smiciklas2011}. }
    \label{tab:my_label-1}
\end{table}

For several resonant mode haloscopes, we pay particular attention to their bandwidth. The bandwidth of NMR haloscopes depends on the relaxation of NG nuclear spins. To cover a given ALP mass range, the number of measurement of NMR magnetometers equals the corresponding frequency range divided by the bandwidth of haloscopes. Under hybrid spin resonance regime, the bandwidth of NG nuclear spins is broadened by strongly coupled AM electron spins, which is related to  the relaxation of electron spins.

Spin-exchange interaction between noble-gas nuclear spins and alkali-metal electron spins provides an in-situ noble-gas polarization detection method, significantly improving the response of ALPs.  Although haloscopes with alkali-noble-gas have presented striking sensitivity to nonstandard-model spin interactions, the achieved sensitivity still falls short of the theoretical limit. All of these alkali-noble-gas haloscopes mentioned above involve AM and NG atoms, optical pumping, magnetization detection, magnetic field shield and manipulation. Instability in any of these components can lead to measurement errors. In particular, the magnetic noise generated by the external environment or system materials, rotation noise, and vibration noise may cause false effects. Further improvement of the performance requires considering the potential aspects as follows:

\textit{Optical pumping:\,} Fluctuations in the light intensity, frequency, alignment and polarization of pump light have a significant effect on the spin polarizations \cite{Sato2018}. Some closed-loop control methods, such as using an acousto-optic modulator \cite{Kwee2012}, an electro-optic modulator \cite{LIU2013}, or a liquid crystal variable retarder \cite{pan2005integrated}, have been used to reduce power fluctuation. Frequency stabilization using the saturation absorption spectrum \cite{tang1974laser} provides a relatively stable condition. However, the effects of pump light fluctuations cannot be completely circumvented. Especially, the low-frequency fluctuation of light alignment is hard to eliminate. Customized pump modes, such as pulsed pumping \cite{Lee2019, Wang2020,Wang2022New,Hunter2022}, should be explored. The improvement of coherence time, polarization, and uniformity of NG nuclear spins is a universal goal in these haloscopes \cite{Wei2023, Gemmel2010PRD, Klinger2023}. For the SC comagnetometer, higher NG nuclear polarization can help the system work at higher frequencies to avoid various low-frequency noises. Although hybrid SEOP has already been proven to be effective, the density ratio of hybrid alkali species, cell geometry and pump light intensity distribution should be optimized for each configuration. More attention should be paid to the relaxation mechanisms of NG nuclear spins due to the gradient Fermi-contact interactions, the interaction between different NG species, and the usual gradient magnetic field. 

\textit{Magnetization detection:\,} An in-situ atom magnetometer carries out the nuclear magnetization through the optical rotation of a linearly polarized probe light \cite{Wu1986,Opechowski1953,Heng2023}. The methods to suppress direct impacts of probe light fluctuations are similar to those for pump light. The probe light is modulated by a Faraday modulator \cite{Allred2002} or a photoelastic modulator \cite{Vasilakis2009} to reduce low-frequency noise. The multipass cell probe can be used to further improve the signal-noise ratio \cite{Zhang2023,Jiang2021}. Apart from that, with the improvement of sensitivity, photon shot noise gradually becomes apparent. The use of squeezed states is proposed to overcome photon shot noise limitation \cite{Vittorio2004,Wolfgramm2010,Malnou2019,Irastorza2021,Wu2023Quantum}. SQUID magnetometer is also a powerful tool to measure the magnetization of noble-gas nuclei to overcome the difficulties of optical rotation detection.

\textit{Magnetic field shield and compensation:\,} Precise magnetic field manipulation \cite{Sachdeva2019} and low magnetic noise are essential for improving sensitivity \cite{Bloch2022SA}. Earth's (or laboratory) magnetic field is the main disturbance for pseudomagnetic field exploration. A multilayer $\mu$-metal magnetic shield can provide a high magnetic shielding factor $\sim 10^6$ \cite{Allred2002, Wei2020}. The magnetic noise from the environment and shield itself should be suppressed using Ferrite,  nanocrystalline or superconductor materials \cite{xu2023constraining}. With five layers of $\mu$-metal shields and an inner Ferrite shield, the single channel and the differential-mode magnetic noise are evaluated as 0.7 fT/Hz$^{1/2}$ and 0.1 fT/Hz$^{1/2}\,@30$\,Hz \cite{ma2022Analysis}. The electric heater should also be optimized to reduce the magnetic noise from the heating current inside the magnetic shield. Meanwhile, highly precise uniform and gradient magnetic field coils are indispensable for atom precession \cite{Sakamoto2015, Chen2023}. The design of coils using global optimization algorithms is an attractive approach \cite{Wang2022, Wang2023Design}. 

\textit{Atom ensembles:\,} NG nuclear spins act as the sensitive source, while AM electron spins serve to polarize NG nuclear spins and read out the magnetization. The alkali-metal-noble-gas pairs, which involve different Fermi-contact interactions and gyromagnetic ratios, can be optimized for the best capacity \cite{Ghosh2010,Su2021}. The atomic number density determines relaxations (e.g., spin exchange relaxation, spin destruction relaxation, and spin diffusion relaxation) and affects the equilibrium polarization \cite{Rosenberry2001}. Thus, maintaining the stability of the cell temperature, which corresponds with the number density, is essential for accurate measurements.

\textit{Others:\,} In addition to these aspects, Earth’s rotation as a directional angular velocity will cause the frequency shift of NG nuclear spins. The fluctuation of the angle between the Earth's rotational axis and the haloscope's sensitive axis disturbs the exploration. A ring laser gyroscope is helpful for identifying the Earth’s rotation orientation \cite{Terrano2021, Zhang2023}. The longitudinal interactions between the nuclei or from the AM effective field cause the frequency drifts in CC comagnetometers. Precise state initialization, transversely pumping, decoupling sequences, and RF pulses are ways to realize no net longitudinal polarization \cite{Limes2018,Limes2019, Korver2015Synchronous}. However, the drifts induced by pulses themselves should be further decoupled \cite{Terrano2021}. Vibration noise from mechanical vibration, human activity, sound, air convection or tidal waves is prevalent in these haloscopes, which can be partly suppressed by both active and passive vibration isolation \cite{Hensley1999, Yao2021, LIU2023Modeling, Liu2023Investigation}. Novel operation regimes, such as exceptional points of the non-Hermitian system \cite{Zhang2023Stable,Tang2023PT}, are expected to bring hope.

The continuous exploration of novelty atomic manipulation methods is expected to break the conventional noise limits. Researches in this area are still advancing, and ALP explorations are expected to be constantly improved.



\section*{CONFLICTS OF INTEREST}
The authors declare that there are no conflicts of interest regarding the publication of this paper.
\section*{Acknowledgments}

We would like to thank Dmitry Budker for the helpful discussions and paper revision. The work of KW  is supported by NSFC under Grants No. 62203030 and 61925301 for Distinguished Young Scholars, and by the Innovation Program for Quantum Science and Technology under Grant 2021ZD0300401. The work of WJ  is supported by the DFG Project ID 390831469: EXC 2118 (PRISMA+ Cluster of Excellence), by the German Federal Ministry of Education and Research (BMBF) within the Quantumtechnologien program (Grant No. 13N15064), by the COST Action within the project COSMIC WISPers (Grant No. CA21106), and by the QuantERA project LEMAQUME (DFG Project No. 500314265).  The work of JL is supported by NSFC under Grant No. 12075005, 12235001.


\end{document}